\documentclass[pdflatex,sn-nature]{sn-jnl}
\usepackage{graphicx}%
\usepackage{multirow}%
\usepackage{amsmath,amssymb,amsfonts}%
\usepackage{amsthm}%
\usepackage{mathrsfs}%
\usepackage[title]{appendix}%
\usepackage{xcolor}%
\usepackage{textcomp}%
\usepackage{manyfoot}%
\usepackage{booktabs}%
\usepackage{algorithm}%
\usepackage{algorithmicx}%
\usepackage{algpseudocode}%
\usepackage{listings}%
\usepackage{lmodern}
\usepackage{anyfontsize}
\usepackage{xr}

\begin{document}

    \title[Kinetically Trapped Nanocrystals with Symmetry-Preserving Shapes]{Kinetically Trapped Nanocrystals with Symmetry-Preserving Shapes}

\author*[1]{\fnm{Carlos L.} \sur{Bassani}}\email{carlos.bassani@fau.de}

\author*[1]{\fnm{Michael} \sur{Engel}}\email{michael.engel@fau.de}

\affil[1]{\orgdiv{Institute for Multiscale Simulation}, \orgname{Friedrich-Alexander-Universität Erlangen-Nürnberg}, \orgaddress{\street{Cauerstrasse 3}, \city{Erlangen}, \postcode{91058}, \country{Germany}}}

\abstract{The shape of nanocrystals is crucial in determining their surface area, reactivity, optical properties, mechanical strength, and self-assembly behavior. Traditionally, shape control has been achieved through empirical methods, highlighting the need for a more refined theoretical framework. A comprehensive model should account for the kinetic factors at distinct stages of the shape-formation process to identify the key determinants of nanocrystal morphology. By modulating kinetics at terraces, ledges, and kinks, we reveal that the primary factors are the adatom nucleation energies and the geometry of growth islands. Transient sites dominate the growth process, leading to kinetically trapped, metastable shapes. We illustrate these concepts with face-centered cubic nanocrystals, demonstrating diverse shape evolutions, including surface roughening and the preservation of crystal symmetry in cubes, octahedra, rhombic dodecahedra, and their truncated variants. This study reveals the mechanisms driving the formation of cubic nanocrystal shapes and offers guidance for their precise synthesis.}



\maketitle

Precise control of the external shape of nanocrystals (NCs) is essential for the development of advanced nanomaterials \cite{Grzelczak2010, Boles2016}. Various shapes were achieved in the synthesis of noble metal NCs, commonly exhibiting a face-centered cubic (fcc) crystal structure. Shapes can preserve the symmetry of the underlying crystal structure, such as cubes, octahedra, cuboctahedra \cite{Xia2012, Xia2012_cubesToOctahedra}, and rhombic dodecahedra \cite{AhnLeeNam2013}, or spontaneously break symmetry, such as tetrahedra \cite{Sun2021, Langille2012}, bitetrahedra, decahedra, icosahedra \cite{Langille2012, Ma2020}, nanorods \cite{PerezJuste2005, YeCaglayan2012}, and nanoplates \cite{WangKang2014, ChoiKim2023}. 
Further shape complexities include branching \cite{Liberato2003, ChenWang2003}, concave and high Miller index facets \cite{Zhou2009}, and chiral features\cite{Googasian2022, Lee2018}. Diverse NC shapes enable localized surface plasmon resonances of importance in catalysis, spectroscopy, and sensing \cite{Willets2007, Langer2019}, and their assembly into colloidal crystals with diverse functionalities \cite{Solomon2007, Bassani2024_KITP}.

The synthesis of NCs typically occurs in a solution containing a salt of the crystal component, with the noble metal ion referred to as the precursor \cite{BaiOuyang2019, Nguyen2022}. Ions and ligands act as facet-selective capping agents \cite{Yang2020} or shape-directing agents \cite{Balankura2016}, modulating the growth velocity of different crystallographic directions and leading to the formation of different shapes \cite{Fichthorn2016}. Competitive kinetic and equilibration effects during growth can be tuned by the type and concentration of precursors, ions, and ligands, and by the pH, temperature, and cooling rate. While the synthesis conditions for different NC shapes are fairly well documented, the understanding of their formation mechanisms remains incomplete. Determining them through atomistic simulations is challenging due to their multiscale nature, covering quantum effects of metallic and metal-organic bonding \cite{Daw1993, Fichthorn2014, ZhouFichthorn2014}, enthalpic and entropic interactions in crystallization/self-assembly \cite{VoGlotzer2022}, long-range interactions due to the presence of ions and water \cite{Jorgensen1983, Abascal2005}, the descriptions of long-molecules (ligands) and their adsorption to NC facets \cite{Balankura2016}, and the influence of larger scales in environments that present uneven concentration or temperature fields \cite{Bassani2020, Ortellado2020}.

Simulating NC shape formation using potential energy models fitted with density functional theory to feed molecular dynamics (MD) simulations is computationally prohibitive due to NC size spanning tens of nanometers encompassing millions of atoms. In contrast, geometric construction models, known as Wulff shapes \cite{Marks2016}, provide an accessible, yet simplistic method to predict NC shapes. A link between fundamental knowledge --the periodic table and potential energy models-- and growth velocities, which could extend classical Wulff constructions, does not exist. Current knowledge remains system- and composition-dependent and has been derived from atomistic simulations capturing only parts of the NC formation process \cite{Broughton1982, LiangKusalik2010, Bassani2024_JCP}.

In this study, we bridge energy, growth velocity, terrace-ledge-kink (TLK) crystallization theory \cite{Jeong1999}, and the emergence of shapes with kinetic Monte Carlo simulations that achieve sufficient system sizes and extended time scales with atomistic resolution. We hypothesize that a small set of key determinants derived from the energy model, representing the combined effects of precursors, ions, ligands, and solvents, guides the growth of NCs into various shapes. We explain when and why NCs composed of the same material and crystal structure exhibit different shapes that are kinetically trapped and correspond to metastable equilibria.

\section*{Nanocrystal Growth Simulations}

\begin{figure}[b]
    \centering
    \includegraphics[width=1.0\textwidth]{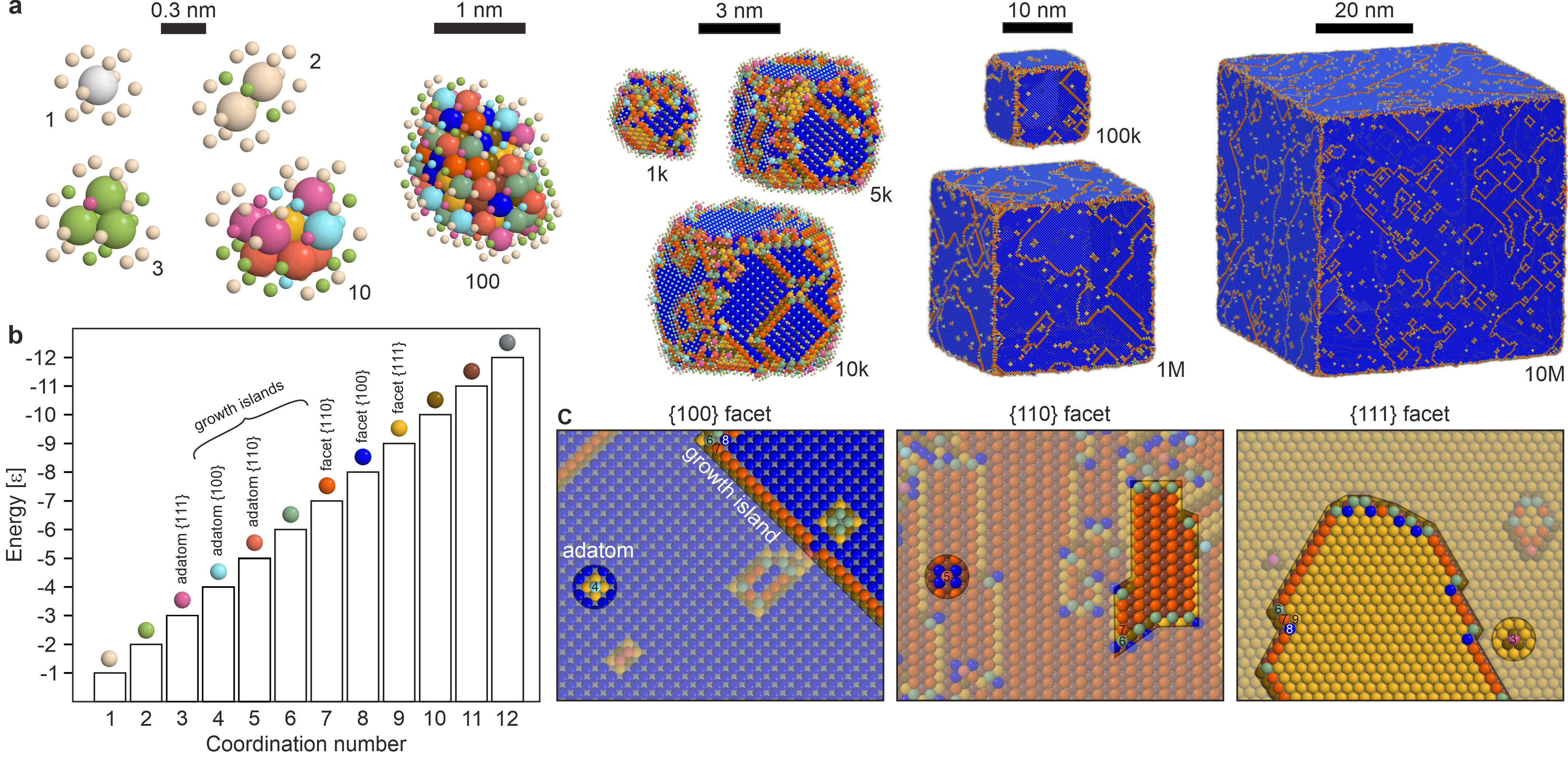}
    \caption{\textbf{Rejection-free kinetic Monte Carlo simulation to predict NC shape.} (a)~Exemplary growth trajectory of fcc NC into a cube. Large spheres represent atoms and small spheres represent growth sites. The inset numbers represent the number of atoms composing the NC. The scale bars are for a lattice parameter of 0.4~nm, representative of noble metals. (b)~Energy scales with coordination number. Coordination numbers relate to NC surface topology features, including adatoms, growth islands, facets, and bulk atoms. (c)~Exemplary adatoms and growth islands in primary facets. Number insets represent coordination numbers.
    }
    \label{fig:Method}
\end{figure}

We simulate NC shape formation using a rejection-free kinetic Monte Carlo (rfKMC) method, which bridges the gap between MD simulations and geometric construction models. In this approach, growth sites for adding atoms are dynamically positioned around the NC surface and continuously updated as the NC evolves (Fig.~\ref{fig:Method}a). The simulations scale up efficiently, achieving a system size that compares to NC synthesis, in the scale of tens of nanometers. Two Monte Carlo moves are considered. Etching moves remove a surface atom $i$ at an etching rate $r_{\text{e},i}$. Growth moves add an atom at a growth site $i$ with a growth rate $r_{\text{g},i}$. The growth rate (negative sign) and etching rate (positive sign) are given by
\begin{equation} \label{eq:r}
    r_{i} \propto \exp \left(\mp \frac{E_i}{kT} \right),
\end{equation}
where $kT$ is the Boltzmann constant multiplied by the temperature, and the energy $E_i$ represents the (free) energy change of adding an atom at the growth site $i$. We assume this energy depends linearly on the coordination number $z_i$ as depicted in Fig.~\ref{fig:Method}b. In the case of an fcc lattice, $E_i$ is discrete with twelve energy levels given by $E_i = -z_i \epsilon$, where $1\leq z_i\leq 12$ and $\epsilon$ is the bond energy, with dimensionless counterpart $\epsilon^{*} = \frac{\epsilon}{kT}$.

The simulation starts with a single atom and twelve equivalent growth sites. As atoms are added (sequence in Fig.~\ref{fig:Method}a), new growth sites are created, and the energy of existing growth sites is updated. Through repeated iterations, the NC shape gradually forms and evolves. We implemented the rfKMC method using a cell list to expedite the calculation of coordination numbers. Growth and etching rates are stored in a binary decision tree, branched by a kinetic parameter $p$ (Supplementary Fig.~S1) that represents the fraction of growth moves, ranging from $p=1$ (irreversible growth) to $p=0$ (irreversible etching). When $p = 0.5$ (equilibration), growth and etching events are equally probable, allowing high-energy atoms to etch and regrow at low-energy sites, mimicking surface diffusion. With an efficient implementation of the rfKMC method, we can simulate the growth of NCs on the scale of tens of nanometers within minutes of computation time. The Method section further discusses the implementation of the rfKMC method.

\section*{Shape Diagrams}

\begin{figure}[b]
    \centering
    \includegraphics[width=1.0\textwidth]{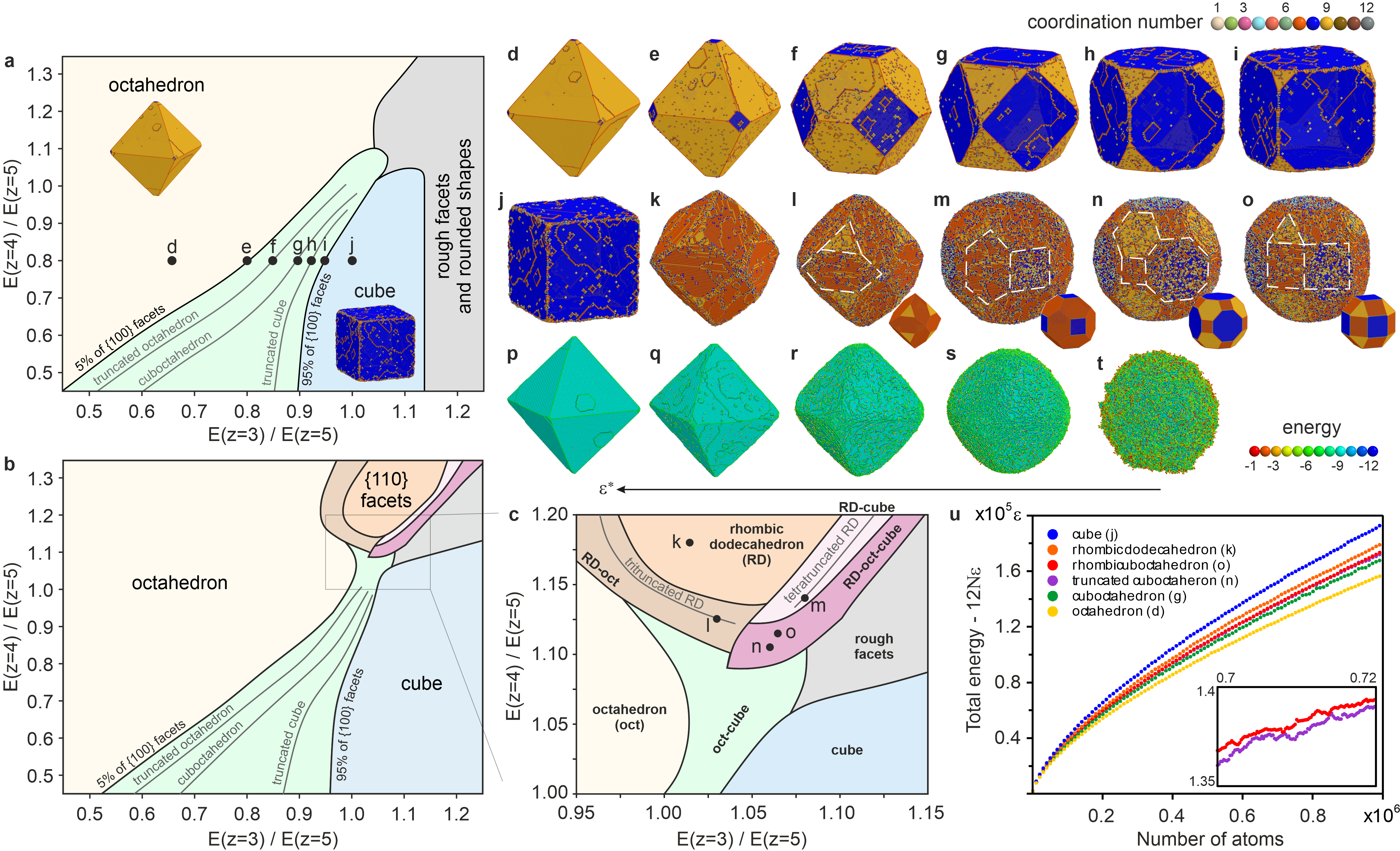}
    \caption{\textbf{Kinetically trapped fcc NCs with symmetry-preserving shapes.} Shape diagrams for (a)~irreversible ($p=1$) and (b,c)~reversible ($p=0.6$) growth. The axes of the shape diagram are energy ratios associated with coordination numbers 3, 4, and 5, which control adatom nucleation on primary facets. Simulation snapshots for (d)~octahedron, (e)~octahedron with truncated $\{100\}$ tips covering 5\% of the surface area, (f)~truncated octahedron, (g)~cuboctahedron, (h)~truncated cube, (i)~cube with truncated $\{111\}$ tips covering 5\% of the surface area, (j)~cube, (k)~rhombic dodecahedron, (l)~tritruncated rhombic dodecahedron, (m)~tetratruncated rhombic dodecahedron, (n)~truncated cuboctahedron, and (o)~rhombicuboctahedron. (p-t)~Sensitivity of shape to ratio of bond energy and temperature $\epsilon^{*} $, showing (p)~smooth ($\epsilon^* = 5$) and (q)~roughened facets ($\epsilon^* = 2.5$), (r)~rounded tips and edges ($\epsilon^* = 1.25$), (s)~isotropic growth leading to spheres ($\epsilon^* = 0.83$), and (t)~pseudo-dendritic morphology ($\epsilon^* = 0.33$). (u)~Evolution of total potential energy of various kinetically trapped NC shapes. NCs grown to 1M atoms. Shape diagrams of (a-c) for $\epsilon^* = 5$, and of (b,c) further consider $\frac{E(z=6)}{E(z=5)} = 1.1$. Video~S1 shows growth trajectories and Supplementary Table~S1 the simulation input parameters. 
    }
    \label{fig:ShapeDiagrams}
\end{figure}

We aim to identify the minimum requirements to form diverse polyhedral NC shapes. Fig.~\ref{fig:Method}b,c show atoms belonging to different surface features and their coordination number. Atoms at the primary facets have coordination numbers $z_{\{110\}} = 7$, $z_{\{100\}} = 8$, and $z_{\{111\}} = 9$. Consequently, the surface energies of these facets are $E_{\{111\}} < E_{\{100\}} < E_{\{110\}}$, resulting in the equilibrium Wulff shape \cite{Marks2016} of fcc being an octahedron. One might expect that lowering the energy level associated with the coordination number of $\{100\}$ facets would result in cubes. However, this is not the case; the NC shape remains an octahedron. Thus, it is not the energy of atoms within facets that control the NC shape, but rather the growth pathway that kinetically traps NC shapes into metastable equilibrium.

Shape-directing agents, such as ligands and ions, can modify the growth velocity of crystal facets by capping certain crystallographic directions \cite{Xia2012}, or by affecting the hole-opening barriers in the potential of mean force, influencing diffusion of precursors \cite{Fichthorn2016}. These kinetic effects occur during facet formation, thus the energy levels associated with lower coordination numbers, $z < 7$, are crucial for shape selection. According to TLK crystallization theory \cite{Jeong1999}, the growth-limiting step is adatom nucleation on a smooth facet. They occur at coordination numbers $z^\text{ad}_{\{111\}} = 3$, $z^\text{ad}_{\{100\}} = 4$, and $z^\text{ad}_{\{110\}} = 5$ for the primary facets (Fig.~\ref{fig:Method}b,c), which predicts growth velocities $v_{\{111\}} < v_{\{100\}} < v_{\{110\}}$. Following the kinetic Wulff shape theory \cite{Fichthorn2016}, facets that grow faster disappear, leaving the slow-growing facets to define the NC shape. Consequently, this theory predicts a shape bounded by $\{111\}$ facets, that is, an octahedron.

Both equilibrium and kinetic Wulff shape theories yield similar results thus far. To bias the formation of other shapes, we must adjust the growth rates (the energy) associated with the coordination numbers $3\leq z\leq 5$ that represent adatom nucleation. Using the energy level $E(z=5)$ as a reference, we summarize the effect of the ratios $\frac{E(z=3)}{E(z=5)}$ and $\frac{E(z=4)}{E(z=5)}$ on NC shape formation in two shape diagrams, for irreversible growth ($p=1$, Fig.~\ref{fig:ShapeDiagrams}a) and for growth closer to equilibrium ($p = 0.6$, Fig.~\ref{fig:ShapeDiagrams}b,c). The coexistence of kinetic and equilibration effects when growth is reversible enables the formation of $\{110\}$ facets for the first time. Supplementary Fig.~S2 shows the role of growth reversibility (the kinetic parameter $p$) in shape transformations. The transition lines of the shape diagrams are based on the surface coverage of facets, and their definition is further discussed in the Method section. 

The shape diagrams capture the formation of shapes spanning the entire family of symmetry-preserving polyhedra truncated by primary facets, namely cubes, octahedra, rhombic dodecahedra, and their various truncations (Fig.~\ref{fig:ShapeDiagrams}d-o). Notably, several of these truncated shapes are Archimedean solids, such as the truncated cube, the truncated octahedron, the cuboctahedron, the truncated cuboctahedron, and the rhombicuboctahedron. Equilateral chamfered solids also appear, such as the tritruncated rhombic dodecahedron (equilateral chamfered octahedron) and the tetratruncated rhombic dodecahedron (equilateral chamfered cube). Video~S1 shows the growth trajectory of these NC shapes from a single atom, and Supplementary Table~S1 compiles the simulation input parameters. Shapes with a single type of facet occupy two-dimensional regions of the parameter space. In contrast, Archimedean and equilateral chamfered solids with two types of facets with defined surface area ratios occur along coexistence lines. Shapes with three types of facets with defined surface area ratios occur at a single point. The dimensionality of these regions mirrors the Gibbs phase rule in phase diagrams.

We emphasize that only the energy of growth sites with low coordination numbers ($3\leq z\leq 6$) had to be changed to realize diverse polyhedral NC shapes. While these coordination numbers temporarily guide growth, they are transient. The fact that the energy of atoms with higher coordination numbers, representing facet and bulk atoms, can be kept constant demonstrates that the final NC shapes are kinetically trapped in metastable equilibrium. Fig.~\ref{fig:ShapeDiagrams}u tracks the total energy of NCs as they grow into different shapes, indicating once more that the octahedron shape constitutes the lowest energy. The energy plotted in Fig.~\ref{fig:ShapeDiagrams}u is rescaled to a reference of $12N\epsilon$, which represents a crystal formed of $N$ bulk atoms $(z=12)$, each with bond energy $\epsilon$. The inset of Fig.~\ref{fig:ShapeDiagrams}u shows that fluctuations in energy occur due to partially grown facets. Supplementary Fig.~S3 provides more details about energy evolution. Once the NC grows to a certain shape, changing the kinetic parameter to equilibrium growth ($p = 0.5$) does not transform the shape. Instead, equilibration relaxes the surface and truncates tips and edges (Video~S1). Such equilibration is similar to suddenly changing the solution conditions during NC synthesis to stop growth. Metastable shapes found in experiments remain stable for extended periods of time, permitting applications of NC with shapes beyond the unique equilibrium shape.

\section*{Facet Roughening and Isotropic Growth}

Fig.~\ref{fig:ShapeDiagrams}p-t illustrates the impact of varying bond energy $\epsilon^*$ on NC morphology. At high values of $\epsilon^*$, NCs exhibit faceted shapes. As $\epsilon^*$ decreases, the facets start to roughen, leading to the development of rounded tips and edges, and ultimately resulting in spherical shapes. In extreme cases, pseudo-dendritic surface features can emerge. Note that the rfKMC method is not suitable for this limit, as dendritic growth evolves from mass transfer-limited crystallization \cite{Ortellado2020}, and the simulations do not account for the concentration field in the solution phase.

When $\epsilon^*$ is large, the energy of growth sites on the NC surface differ significantly. Consequently, atoms are preferentially added to growth sites with higher coordination numbers (lower energies), following TLK crystallization theory \cite{Jeong1999}. This preference results in a layer-by-layer growth mode. The simulations mimic atom diffusion (hopping) on the NC surface over short distances until they find the optimal site for attaching. As $\epsilon^*$ decreases, the growth rates of sites with different coordination numbers become more similar. Therefore, the growth site where an atom settles may not minimize energy globally but only locally within its diffusion range. This leads to the formation of multiple growth islands on the NC surface, causing surface roughening. When the diffusion length is such that atoms cannot hop, and instead they attach to the first site they encounter, the NC grows isotropically, forming a spherical shape.

\section*{From Energy to Growth Velocity}

\begin{figure}[b]
    \centering
    \includegraphics[width=1.0\textwidth]{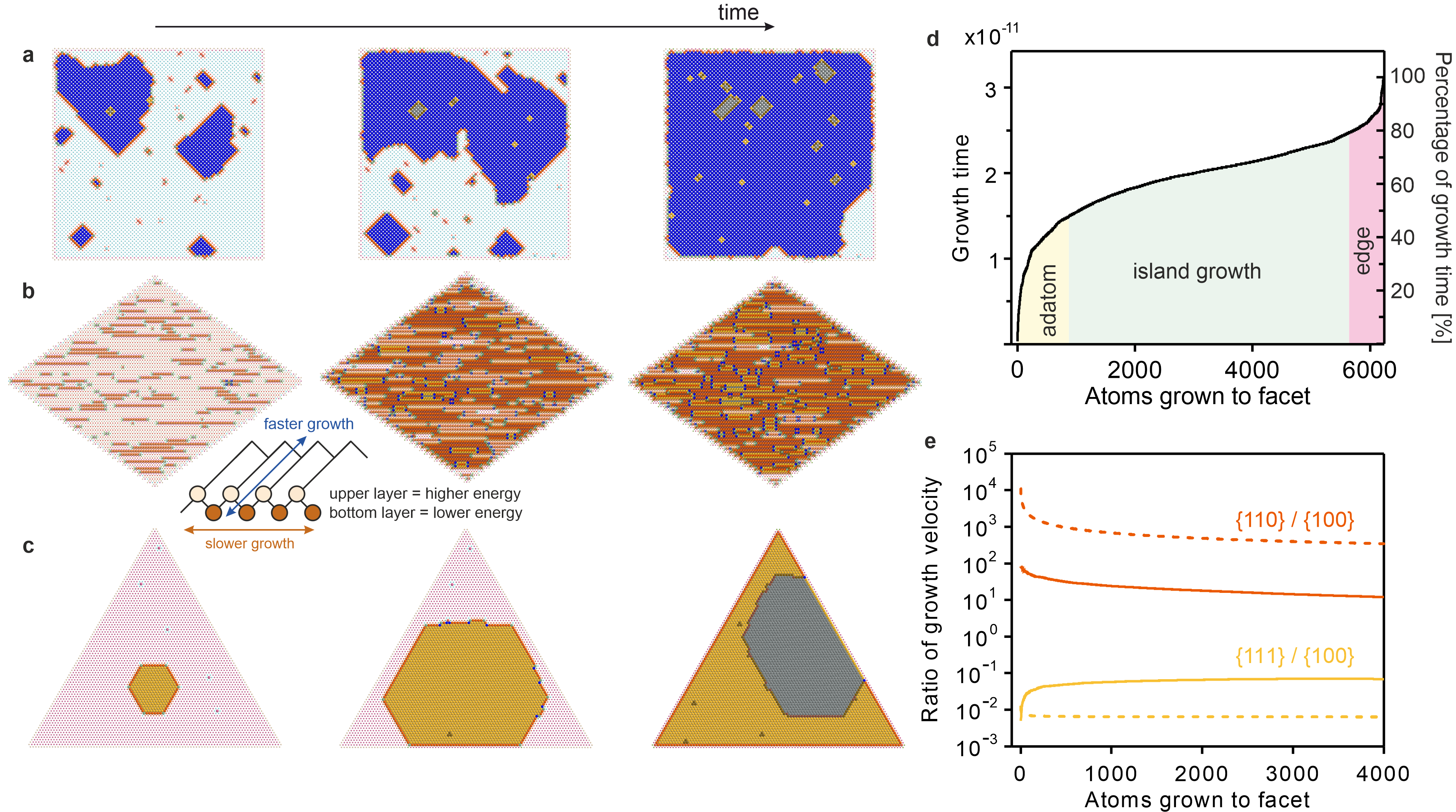}
    \caption{\textbf{Estimation of growth velocity.} (a-c)~Snapshots showing islands growing on (a)~$\{100\}$, (b)~$\{110\}$, and (c)~$\{111\}$ facets. Video~S2 shows growth trajectories. Inset shows preferential growth parallel to the grooves of the zigzag pattern of $\{110\}$ that explains the formation of elongated (linear) islands. (d)~Exemplary growth time of a facet, showing three stages of growth, namely adatom nucleation leading to island formation (yellow), island growth (green), and edge effects (red). (e)~Ratios of growth velocities between primary facets for the analytic model (dashed lines) and rfKMC simulations (continuous lines). Simulations of rfKMC over primary facets for $\frac{E(z=3)}{E(z=5)} = 0.6$, $\frac{E(z=4)}{E(z=5)} = 0.8$, $\epsilon^* = 5$, and $p = 1$.
    }
    \label{fig:GrowthVelocity}
\end{figure}

We now turn our attention to the relationship between energy (a parameter of the atomic scale, fundamentally described by the periodic table) with growth velocity (of importance to the engineering of NCs with various shapes). We explore the importance of the scale between atoms and NCs, namely the growth islands composed of thousands of atoms forming over the NC facets. We performed rfKMC simulations over planar 2D films to estimate growth velocities of primary facets, Fig.~\ref{fig:GrowthVelocity}a-c and Video~S2. Supplementary Fig.~S4 shows the evolution of the growth rate of a facet, that is, the sum of the growth rate of all growth sites composing the facet, known as the Rosenbluth factor. Atomistic fluctuations due to the filling of surface and ledge vacancies are averaged by integration when estimating growth time in Fig.~\ref{fig:GrowthVelocity}d.

There are three stages of growth: Adatom nucleation (yellow) until reaching an island of critical size is the growth-limiting step and accounts for approximately 12\% of the atoms grown to the facet in 50\% of total growth time; Island growth (green) is the faster stage, contributing 80\% of the atoms grown to the facet in just 30\% of total growth time; Edge effects (red) significantly reduce growth velocity and contribute only to 8\% of the atoms grown to the facet while requiring 20\% of total growth time. Supplementary Fig.~S5 compares the growth times on the primary facets under identical conditions. The ratio of growth velocities is inversely proportional to the ratio of growth times and stabilizes during the second stage of growth (Fig.~\ref{fig:GrowthVelocity}e). We aim to determine analytically these growth velocity ratios in steady-state.

For that purpose, we first recognize that the geometry of the growth islands (Fig.~\ref{fig:GrowthVelocity}a-c) depends on the crystallographic direction. We observe rectangular islands on $\{100\}$ facets, linear islands on $\{110\}$ facets, and hexagonal islands on $\{111\}$ facets. Growth on $\{110\}$ facets is unique because it does not preserve the symmetry of the rectangular disposition of atoms in their plane. This is attributed to the non-planar arrangement of surface atoms in a double layer with a zigzag pattern (inset of Fig.~\ref{fig:GrowthVelocity}b). In the bottom layer, atoms have coordination number 11, and growth sites have coordination number 5. In the top layer, atoms have coordination number 7, and growth sites have coordination number 1. The significant difference in coordination numbers requires the bottom layer to form completely before the linear islands in the top layer can merge. Elongated geometries of $\{110\}$ islands also occur when considering the influence of farther neighbors (Supplementary Fig.~S6), with the difference that top and bottom layers can grow concomitantly.

The Method section shows how to use the information gained from Fig.~\ref{fig:GrowthVelocity} to express growth velocities analytically in terms of energy,
\begin{equation} \label{eq:v110perV100}
        \frac{v_{\{110\}}}{v_{\{100\}}} =   a
        \exp \left( \frac{E(z=4) - E(z=5) }{ k T } \right),
\end{equation}
\begin{equation}\label{eq:v111perV100}
    \frac{v_{\{111\}}}{v_{\{100\}}} = b  
    \exp \left( \frac{E(z=4) - E(z=3)}{kT} \right),
\end{equation}
where the exponent terms relate to adatom nucleation, and the parameters $a(E(z=5),E(z=6))$ and $b(E(z=4),E(z=5))$ (see Method section) depend on the energy of growth sites composing the perimeter of growth islands.

\section*{Analytic Prediction of Shapes}

\begin{figure}[b]
    \centering
    \includegraphics[width=1.0\textwidth]{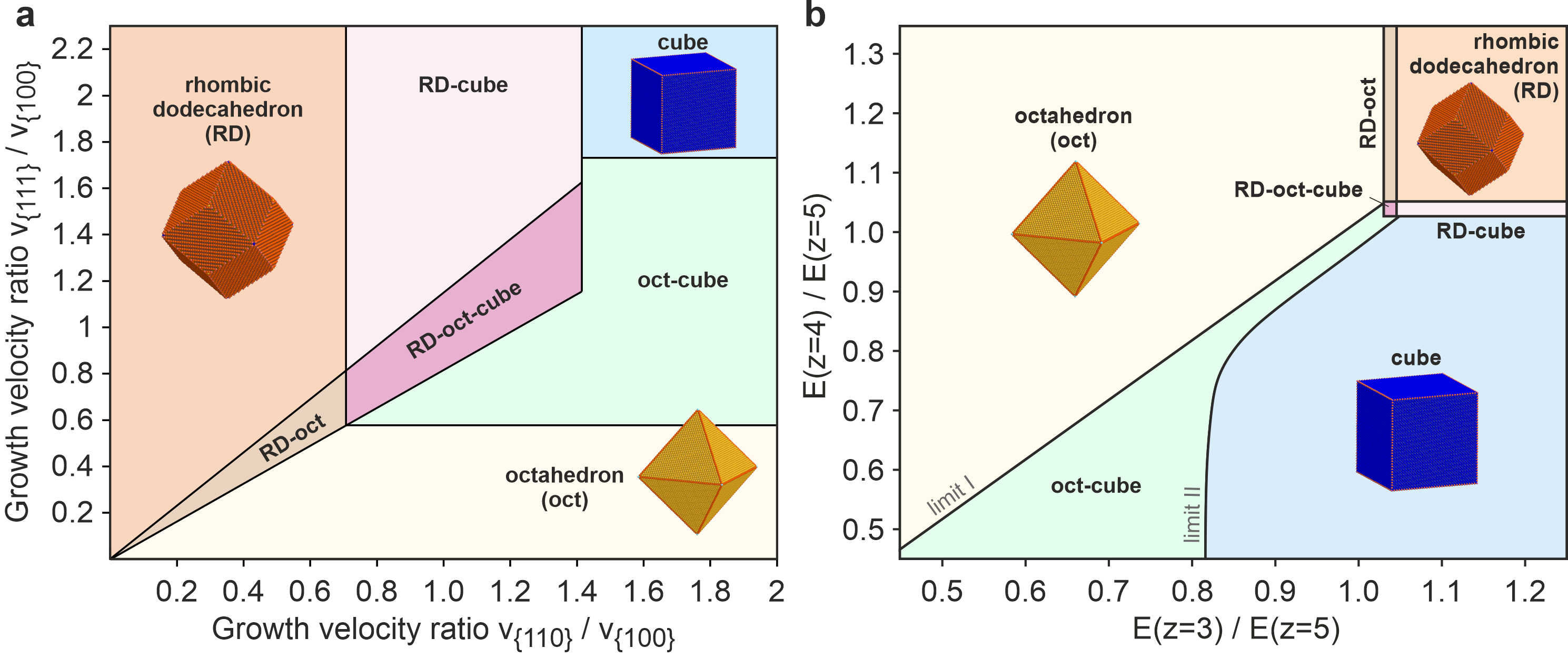}
    \caption{\textbf{Theoretical prediction of NC shapes.} (a)~Shape diagram from the geometric construction of kinetic Wulff shapes based on growth velocity of primary facets (literature knowledge). (b)~Shape diagram based on energy of adatom nucleation in primary facets (current study), derived from the analytic model of Eqs.~(\ref{eq:v110perV100}),(\ref{eq:v111perV100}) combined with the kinetic Wulff shape constructions in (a).
    }
    \label{fig:analyticModel}
\end{figure}

The geometric construction of kinetic Wulff shapes \cite{Fichthorn2016} predicts NC shape analytically from the relationship $\frac{x_{\{hkl\}}}{x_{\{mnp\}}} = \frac{v_{\{hkl\}}}{v_{\{mnp\}}}$, where $x$ is the distance of the facet to the center of the NC. By using this relationship, we construct an analytic shape diagram based on the growth velocity ratios between primary facets (Fig.~\ref{fig:analyticModel}a). Supplementary Fig.~S7 shows the same analytic shape diagram including the ratio of facet surface area and the analytic values of shape transition. With Eqs.~(\ref{eq:v110perV100}),(\ref{eq:v111perV100}), we transform the axes of the analytic shape diagram from growth velocity to energy (Fig.~\ref{fig:analyticModel}b), linking geometric construction of NC shapes with energy in a non-equilibrium approach. This analytic shape diagram reveals good overall agreement with the simulated shape diagrams of Fig.~\ref{fig:ShapeDiagrams} using rfKMC, emphasizing once again that the emergence of kinetically trapped NCs with symmetry-preserving shapes is influenced primarily by the energy/growth rates of adatom nucleation surface sites, and secondarily by ledge growth sites belonging to the perimeter of the growth islands. This explains NC growth, a kinetic process spanning multiple scales, so far considered intractably complex in simple and tangible terms.

The analytic model enlightens individual elements of the shape diagrams of fcc NCs. Limit I of Fig.~\ref{fig:analyticModel}b is oriented along a diagonal of approximately $45^\circ$ and occurs when $\frac{E(z=4)}{E(z=5)} \approx \frac{E(z=3)}{E(z=5)}$, indicating similar rates of adatom nucleation on $\{100\}$ and $\{111\}$ facets. Above the diagonal, faster adatom nucleation on $\{100\}$ facets accelerates growth, eventually leading to the disappearance of these facets and forming octahedra bounded by $\{111\}$ facets. Below this diagonal, $\{100\}$ facets start appearing. The region of truncations between octahedra and cubes is larger than the region of truncations between other shapes. The reason is that adatom nucleation on $\{111\}$ leads to an intermediate state composed of six growth sites of coordination number 4, which competes with adatom nucleation on $\{100\}$ facets. This intermediate state causes the limit II marked in Fig.~\ref{fig:analyticModel}b.

Finally, the analytic model shows that coordination number $z=5$ relates to both adatom nucleation on \{110\} and the growth rate of islands on \{100\} and \{111\}. This explains why the region of rhombic dodecahedra is relatively smaller than the regions of cubes and octahedra. Eq.~(\ref{eq:v110perV100}) further evidences the importance of coordination number 6 in the growth islands of \{110\} facets. Its associated energy level can be biased to enlarge the region of rhombic dodecahedra, as shown in Supplementary~Fig.~S8. Supplementary Fig.~S9 further highlights the impact of the energy associated with coordination number 6 on shape transformations between octahedra and rhombic dodecahedra.

\section*{Links to Chemistry}

The rfKMC simulations and the analytic model in this study are based on simplified assumptions. To design NC shapes, it is essential to link synthesis parameters and fundamental chemistry with the prediction of growth and etching rates at NC surface sites. For example, the NC shape can be hypothesized from the potential energy model describing the element forming the crystal. Using the Lennard-Jones potential for silver \cite{Heinz2008_LJnobleMetals}, the ratios of energy level of adatom nucleation are $\frac{E(z=3)}{E(z=5)} = 0.73$ and $\frac{E(z=4)}{E(z=5)} = 0.83$, which in the shape diagrams of Fig.~\ref{fig:ShapeDiagrams}a-c,\ref{fig:analyticModel}b give an octahedron. The rfKMC simulation using the same Lennard-Jones model indeed forms an octahedron (Supplementary Fig.~S10).

Multiple materials attributes concur to govern the final NC shape. Electron density in metallic bonding, often described using the embedded atom model \cite{Daw1993}, varies considerably near the NC surface. Incorporating solution effects into the modeling process, together with metal-organic bonding of ligands on the NC surface \cite{ZhouFichthorn2014}, further modifies growth rates. The ordering of the ligand shell impacts the diffusion of precursors toward the NC surface and is captured in implicit solvent models predicting a high sensitivity of the hole-opening barrier on the crystallographic direction of the facet \cite{Balankura2016}. The consequence of these attributes is the highly nonlinear dependence of energy levels $E(z)$ on coordination number in our model, including variations of energy levels at the low coordination number sites associated with adatom nucleation and growth islands critical for the diverse shapes of noble metal NCs.

Future studies should integrate rfKMC simulations with existing multiscale modeling frameworks that estimate the potential of mean force from all-atom MD simulations, using energy models based on density functional theory \cite{ZhouFichthorn2014, Fichthorn2014, Fichthorn2016, Balankura2016}. This integration would allow predicting NC shape formation from chemistry alone. For instance, in a solution forming octahedral silver NCs, the presence of polyvinylpyrrolidone (PVP), known to cap $\{100\}$ facets of silver \cite{Yang2020}, induces a shape transformation from octahedra to cubes, passing through their truncations \cite{Xia2012_cubesToOctahedra}, as predicted in our shape diagrams.

\section*{Methods}

\subsection*{Implementation of Rejection-Free Kinetic Monte Carlo Simulations}

The rfKMC method was implemented in the programming language C++. Cell lists with a spacing of one cutoff radius are used to calculate coordination numbers, ensuring linear scaling of computational cost. Each of the twelve coordination numbers in the fcc lattice corresponds to one energy level used to compute the growth/etching rates, as described in Eq.~(\ref{eq:r}). A binary decision tree stores the etching rates of all atoms in the NC surface. Another binary decision tree stores the growth rates of all growth sites above the NC surface.

Supplementary Fig.~S1 illustrates the binary decision tree. Initially, a random number is compared to a user-defined parameter $0 \leq p \leq 1$ to determine whether an atom will grow or etch. Subsequently, another random number decides whether to select the growth site/atom $i$ associated with the node of the binary decision tree, with a probability equal to its growth/etching rate $r_i$, or proceed to the left or right branches of the decision tree, with an accumulated probability based on the rates $r_i$ of the nodes in the respective branches. If the decision is made to move one level down the decision tree, the process is repeated with a new random number.

Sampling of growth and etching events applies only to surface atoms. In the absence of defects (not considered in this study) and porosity (valid for large $\epsilon^*$), the number of stored values scales as $\propto N^{2/3}$. Bulk atoms do not need to be stored in the binary decision tree. This limits the number of operations of the decision tree to $\propto \phi\log(N^{2/3})$ for selecting one rfKMC move (growth or etching), where $\phi$ is a factor depending on the ratio of surface to bulk atoms, that is, on the NC shape. For example, an NC composed of 1M atoms with a spherical shape requires storing only about 50k sites in the tree, and the tree contains about 16 levels.

After selecting the site to grow or etch, the atom is inserted or removed from the simulation domain. The coordination number and energy of all surrounding sites and their growth/etching rates assigned to the decision tree are then updated. Finally, new growth sites are introduced according to the fcc crystal structure.

\subsection*{Definition of Transition Lines in Simulated Shape Diagrams} \label{sec:transitionLinesShapeDiagram}

Facet surface area is determined by counting the number of atoms with specific coordination numbers (7 or 11 for $\{110\}$ facets, 8 for $\{100\}$, and 9 for $\{111\}$) and multiplication with the surface area per atom. Atoms with other coordination numbers are excluded from this analysis. The surface area per atom is calculated as the area of the Wigner-Seitz cell projected onto the $\{hkl\}$ plane, as shown in Supplementary Fig.~S11 and Table~S2.

Supplementary Figs.~S12,13 present contour plots of surface coverage (the ratio $\frac{A_{\{hkl\}}}{A_{\{100\}} + A_{\{110\}} + A_{\{111\}}}$) for different facets $\{hkl\}$. The isolines help identify regions corresponding to octahedra (yellow), cubes (blue), and rhombic dodecahedra (green) in the shape diagrams of Fig.~\ref{fig:ShapeDiagrams}a-c. The exact transition lines, however, require visual inspection due to noise in the surface area estimation This method introduces noise to the surface area estimation when other surface topologies (e.g., tips, edges, growth islands, surface roughness) have the same coordination number as the facets. Equilateral shapes present analytical values of surface coverage, Supplementary Table~S3, which helps identify the gray lines in Fig.~\ref{fig:ShapeDiagrams}a-c of the manuscript.

\subsection*{Analytic Model for Growth Velocity}

The growth rate of an $\{hkl\}$ facet is defined as the sum of the growth rates of all the growth sites within a region $\Gamma$ encompassing the first monoatomic layer over the facet,
\begin{equation} \label{eq:growthRateFacet}
    r_{\{hkl\}}=\sum\limits_{i \in \Gamma} r_{\text{g},i}.
\end{equation}
The time to grow a facet by one atom is inversely proportional to the growth rate of the facet, $t_{\{hkl\}}\propto r_{\{hkl\}}^{-1}$. The exact proportionality constant is not necessary for determining NC shapes, as only the ratios of growth rates and times, not their absolute values, are required. We define $t_{\{hkl\}}(n)$ as the time to grow the $n$-th atom on the facet. Therefore,
\begin{equation} \label{eq:growthTime}
    T_{\{hkl\}}(N) = \sum\limits_{n=1}^N t_{\{hkl\}}(n).
\end{equation}
is the growth time for growing $N$ atoms on the facet $\{hkl\}$.

The time for adatom nucleation, $T^\text{ad}$, is the growth time of the first atom over a smooth facet, that is, for $N=1$. In fcc, adatom nucleation in primary facets depends on coordination numbers $z \in \{3,4,5\}$. Therefore, the ratio of time for adatom nucleation between primary facets is
\begin{eqnarray} \label{eq:tAdatom_100per110}
    \frac{ T_{\{100\} }^\text{ad}}{ T_{\{110\}}^\text{ad}} &=& \exp \left( \frac{E(z=5) - E(z=4) }{ k T } \right),
\\
    \label{eq:tAdatom_100per111}
    \frac{ T_{\{100\}}^\text{ad}}{ T_{\{111\}}^\text{ad}}  &=& \exp \left( \frac{E(z=3)- E(z=4) }{ k T } \right).
\end{eqnarray}
Multiple islands may grow simultaneously on the same facet. On average, one adatom (one new growth island) nucleates every time period $T^\text{ad}$. Therefore, $T_{\{pqm\}}^\text{ad} / T_{\{hkl\}}^\text{ad}$ represents the average ratio of the number of growth islands between facets with Miller indices $\{hkl\}$ and $\{pqm\}$. For example, if $T_{\{111\} }^\text{ad} / T_{\{100\}}^\text{ad} = 2$, then on average twice as many islands grow on $\{100\}$ facets compared to $\{111\}$, assuming facets of similar surface area.

Growth islands on different facets take different times $T^\text{isl}$ to grow, with each island having a growth velocity inversely proportional to this time. Considering multiple growth islands on each facet, the velocity ratio between two distinct crystallographic directions is
\begin{equation} \label{eq:VratioVsTimescalesIslandAndAdatom}
    \frac{ v_{\{hkl\} }}{ v_{\{pqm\} }}  = \frac{ T_{\{pqm\} }^\text{ad}}{ T_{\{hkl\}}^\text{ad}} \frac{ T_{\{pqm\} }^\text{isl}}{ T_{\{hkl\}}^\text{isl}} .
\end{equation}
Considering irreversible growth ($p = 1$) and a layer-by-layer growth mode (high values of $\epsilon^*$), the time to grow one atom is the inverse of the sum of growth rates of all sites $i$ on the ledge sites of the perimeter $P$ of the growth islands. With these assumptions in Eqs.~(\ref{eq:growthRateFacet}),(\ref{eq:growthTime}), the time to grow an island of $N$ atoms from an already nucleated adatom is
\begin{equation} \label{eq:tIslandGeneralSI}
    T_{\{hkl\}}^\text{isl}(N) = \sum_{n=2}^N \left[ \sum_{i\in P} r_{\text{g},i} \right] ^{-1}.
\end{equation}

Based on simulations shown in Figs.~\ref{fig:GrowthVelocity}a-c, we idealize the geometry of growth islands as squares on $\{100\}$, lines on $\{110\}$, and regular hexagons on $\{111\}$ (Supplementary Fig.~S14). We tread $N$ as a continuous variable to simplify calculations, avoiding complexities from non-perfect islands and discrete energy levels. For instance, whereas a growth island on a $\{100\}$ facet with $N=26$ atoms forms a $5 \times 5$ square with one ledge adatom, we approximate the square length to $\sqrt{N} \approx 5.1$. This approximation averages out atomistic fluctuations, irrelevant at the mesoscales of shape formation. We further assume that each atom on the perimeter of the island is associated with a single growth site, ignoring corner sites.

With these simplifications, the time to grow an island on $\{100\}$ and $\{111\}$ facets until $N$ atoms is
\begin{equation} \label{eq:tIsland_100_111}
    T_{\{hkl\}}^\text{isl}(N) = \sum_{n=2}^N  \left[ \phi_{\{hkl\}} \sqrt{n} \exp \left(-\frac{E(z=5)}{k T} \right) \right]^{-1},
\end{equation}
where $\phi_{\{100\}} = 4$ and $\phi_{\{111\}} = 6\sqrt{\frac{2}{3 \sqrt{3}}} \approx 3.72 $ relate to the geometry of the growth island. The expression slightly varies for $\{110\}$ facets given their 1D geometry,
\begin{equation} \label{eq:tIsland_110}
    T_{\{110\}}^\text{isl}(N) = \sum_{n=2}^N \left[ 2 \exp \left( - \frac{E(z=6)}{k T} \right) \right] ^{-1}.
\end{equation}

Eq.~(\ref{eq:tIsland_100_111}) considers a direct transition from adatom nucleation to a growth island modeled with a continuous geometry bounded by growth sites of coordination number 5. The continuous geometry assumption holds for large $N$, but still presents satisfactory results for low $N$ in the purpose of the simplified model we propose. However, of importance is the fact that adatom nucleation on $\{111\}$ generates six growth sites of coordination number 4. Explicitly considering these sites gives
\begin{equation} \label{eq:tIsland_111_modified}
    T_{\{111\}}^\text{isl}(N) = \left[6 \exp \left(-\frac{E(z=4)}{k T} \right) \right]^{-1} +
    \sum_{n=3}^N \left[ 3.72\sqrt{n} \exp \left(- \frac{E(z=5)}{k T} \right) \right]^{-1} .
\end{equation}

Note that coordination number 4 also occurs in all the growth sites of smooth $\{100\}$ facets, but those cannot be explicitly considered in the current simplification hypotheses, given that facet size is not known a priori. We can nevertheless recognize that Eqs.~(\ref{eq:tIsland_100_111}),(\ref{eq:tIsland_111_modified}) limit the value of $T_{\{111\}}^\text{isl}(N)$. These two equations give rise to limits I and II of the analytic shape diagram of Fig.~\ref{fig:analyticModel}b. Finally, by inserting Eqs.~(\ref{eq:tAdatom_100per110})-(\ref{eq:tAdatom_100per111}),(\ref{eq:tIsland_100_111})-(\ref{eq:tIsland_111_modified}) into Eq.~(\ref{eq:VratioVsTimescalesIslandAndAdatom}), we obtain
\begin{equation}
        \frac{v_{\{110\}}}{v_{\{100\}}} =   a
        \exp \left( \frac{E(z=4) - E(z=5) }{ k T } \right),
\end{equation}
\begin{equation}
    \frac{v_{\{111\}}}{v_{\{100\}}} = b  
    \exp \left( \frac{E(z=4) - E(z=3)}{kT} \right),
\end{equation}
where $b \in \{b_\text{I}, b_\text{II}\}$ indicates limit I and II, with $b_\text{I} = 0.93$ and
\begin{equation}
    a = \frac{ \sum_{n=2}^N \left[ 4 \sqrt{n} \exp \left(- \frac{E(z=5) }{ k T } \right) \right] ^{-1} }
        { \sum_{n=2}^{N} \left[ 2  \exp \left(- \frac{E(z=6)}{k T} \right) \right] ^{-1} },
\end{equation}
\begin{equation}
    b_\text{II} = 
    \frac{\sum_{n=2}^N  \left[ 4 \sqrt{n} \exp \left(-\frac{E(z=5)}{k T} \right) \right]^{-1} }
    {\left[6 \exp \left(- \frac{E(z=4)}{k T} \right) \right]^{-1} +
    \sum_{n=3}^N \left[ 3.72 \sqrt{n} \exp \left(- \frac{E(z=5)}{k T} \right) \right]^{-1} }.
\end{equation}

\subsection*{Geometric Construction Models for Shape Prediction}

Geometric construction models provide an accessible and comprehensible method for predicting NC shapes. The classic approach, equilibrium Wulff shapes, estimates the relative sizes of facets by minimizing surface energy \cite{Wulff1901, Marks2016}. The distance of a facet to the center of the NC, $x_{\{hkl\}}$, is related to the surface energy $\sigma$ of each facet by $\frac{x_{\{hkl\}}}{x_{\{mnp\}}} = \frac{\sigma_{\{hkl\}}}{\sigma_{\{mnp\}}}$, where $\{hkl\}$ and $\{mnp\}$ are the Miller indices of two facets. Other equilibrium geometric construction models consider growth from one or two substrates, resulting in Winterbottom \cite{Winterbottom1967} and Summertop \cite{Coninck1993} shapes, or binary component systems, leading to alloy Wulff constructions \cite{Ringe2011}. However, equilibrium models cannot account for kinetic NC growth pathways, which produce NC shapes kinetically trapped in metastable equilibrium.

Kinetic Wulff shapes \cite{Fichthorn2016} offer an improvement over equilibrium Wulff shapes by recognizing that fast-growing facets disappear, leaving only the slow-growing facets to define the NC shape. Energy minimization then influences the growth rate of different facets \cite{Broughton1982, Burke1988}. Kinetic Wulff shapes are derived from the relationship $\frac{x_{\{hkl\}}}{x_{\{mnp\}}} = \frac{v_{\{hkl\}}}{v_{\{mnp\}}}$, where $v$ represents the growth velocity in different crystallographic directions during steady-state growth \cite{Fichthorn2016}.

Fig.~\ref{fig:analyticModel}a (and Supplementary Fig.~S7) uses this relationship to generate a diagram of kinetic Wulff shapes. The truncation between planes is determined analytically, giving rise to the transition values indicated in the diagram. Furthermore, the diagram shows the coexistence lines (gray lines) of equilateral truncations composed of two types of facets, and the point of formation of equilateral truncations composed of three types of facets. The surface area covered by each facet type was determined following the method described above.

\section*{Acknowledgements}
C.L.B. acknowledges a Humboldt Research Fellowship for postdoctoral researchers of the Alexander von Humboldt Foundation and an EAM Starting Grant (EAM-SG23-1) of the Competence Center Engineering of Advanced Materials at Friedrich-Alexander-Universität Erlangen-Nürnberg. M.E. acknowledges support by Deutsche Forschungsgemeinschaft (DFG) under Project-IDs 416229255 (SFB 1411) and 542350250. This research was advanced through discussions during the workshop Nanoparticle Assemblies: A New Form of Matter with Classical Structure and Quantum Function hosted by the Kavli Institute for Theoretical Physics at the University of California, Santa Barbara, supported by the National Science Foundation under Grant No. NSF PHY-1748958. HPC resources provided by the Erlangen National High Performance Computing Center (NHR@FAU) under the NHR project b168dc are gratefully acknowledged. NHR funding is provided by federal and Bavarian state authorities. NHR@FAU hardware is partially funded by DFG under Project-ID 440719683.

\bibliography{references}

\end{document}


\title{Supplementary Material: Kinetically Trapped Nanocrystals with Symmetry-Preserving Shapes}

\author*[1]{\fnm{Carlos L.} \sur{Bassani}}\email{carlos.bassani@fau.de}

\author*[1]{\fnm{Michael} \sur{Engel}}\email{michael.engel@fau.de}

\affil*[1]{\orgdiv{Institute for Multiscale Simulation}, \orgname{Friedrich-Alexander-Universität Erlangen-Nürnberg}, \orgaddress{\street{Cauerstrasse 3}, \city{Erlangen}, \postcode{91058}, \country{Germany}}}

\maketitle

\section*{Description of Videos}

\noindent \textbf{Video S1.} Growth trajectories of NCs grown to 1M atoms, followed by an equilibration ($p = 0.5$). Besides the octahedron, the shapes are kinetically trapped in metastable equilibria, known as kinetic Wulff shapes. The shapes displayed in panels (a-l) were simulated under the same conditions as those shown in Fig.~2 of the manuscript. Table~\ref{table:inputParameters} presents the input parameters for the simulations.

\vspace{12pt}

\noindent \textbf{Video S2.} (a-c)~Growth trajectory over primary facets for same conditions of Fig.~3 of the manuscript. (d)~Growth time vs. the number of atoms grown on the facet. (e)~The growth velocity ratio feeds geometric construction models (Fig.~\ref{fig:WulffConstruction}), which gives the theoretical shape as an octahedron. Simulations with $\frac{E(z=3)}{E(z=5)} = 0.6$, $\frac{E(z=4)}{E(z=5)} = 0.8$, $\epsilon^* = 5$, and $p = 1$.

\begin{figure*}
    \centering
    \includegraphics[width=1.00\textwidth]{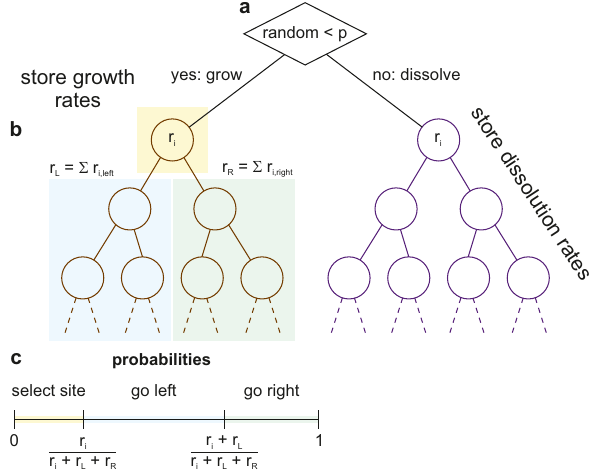}
    \caption{\textbf{Binary decision tree data structure to select sites on the NC surface based on their growth or etching rates.} (a)~A random number between 0 and 1 is compared to the user-defined probability $p$, called kinetic parameter, to determine whether an atom will grow or etch. (b)~The left portion of the binary decision tree stores the growth rates $r_i$ of all growth sites $i$ on the NC surface. The right portion of the decision tree stores the etching rates of all atoms on the NC surface. (c)~At each level of the binary decision tree, a random number determines whether to select the site or to proceed to the left or right branches. This process is repeated through successive levels of the decision tree until a site on the NC surface is selected. This method ensures that a site is selected in every Monte Carlo step of the simulation, making the process rejection-free.
    } \label{fig:binaryDecisionTree}
\end{figure*}

\begin{table}[h]
    \centering
    \caption{\textbf{Input parameters for the rejection-free kinetic Monte Carlo simulations} used to generate the snapshots presented in Fig.~2 of the manuscript, and Fig.~\ref{fig:kineticallyTrapped}, as well as the trajectories in Video~1. All other energy levels are estimated as $z_i \epsilon^*$, where $z_i$ is the coordination number.}
    \label{table:inputParameters}
    
    \begin{tabular}{cclcccccc} 
        \toprule
        Fig.
        & Video
        & Shape
        & \makecell{Number \\of atoms}
        & $\epsilon^*$
        & $p$
        & $\frac{E(z=3)}{E(z=5)})$
        & $\frac{E(z=4)}{E(z=5)})$
        & $\frac{E(z=6)}{E(z=5)})$
        \\
        \midrule

        2d, \ref{fig:kineticallyTrapped}a
        & 1a
        & Octahedron
        & $10^6$
        & 5
        & 1
        & 0.66
        & 0.80
        & 1.2
        \\

        2e
        & 1b
        & \makecell[tl]{Octahedron with \\5\% of \{100\} tips truncated}
        & $10^6$
        & 5
        & 1
        & 0.80
        & 0.80
        & 1.2
        \\

        2f
        & 1c
        & Truncated octahedron
        & $10^6$
        & 5
        & 1
        & 0.86
        & 0.80
        & 1.2
        \\

        2g, \ref{fig:kineticallyTrapped}b
        & 1g
        & Cuboctahedron
        & $10^6$
        & 5
        & 1
        & 0.90
        & 0.80
        & 1.2
        \\

        2h
        & 1d
        & Truncated cube
        & $10^6$
        & 5
        & 1
        & 0.92
        & 0.80
        & 1.2
        \\

        2i
        & 1h
        & \makecell[tl]{Cube with \\5\% of \{111\} tips truncated }
        & $10^6$
        & 5
        & 1
        & 0.94
        & 0.80
        & 1.2
        \\

        2j, \ref{fig:kineticallyTrapped}f
        & 1l
        & Cube
        & $10^6$
        & 5
        & 1
        & 1.00
        & 0.80
        & 1.2
        \\

        2k, \ref{fig:kineticallyTrapped}e
        & 1i
        & Rhombic dodecahedron
        & $10^6$
        & 5
        & 0.6
        & 1.015
        & 1.180
        & 1.1
        \\

        2l
        & 1e
        & \makecell[tl]{Tritruncated \\ rhombic dodecahedron}
        & $10^6$
        & 5
        & 0.6
        & 1.030
        & 1.125
        & 1.1
        \\

        2m
        & 1j
        & \makecell[tl]{Tetratruncated \\ rhombic dodecahedron}
        & $10^6$
        & 5
        & 0.6
        & 1.080
        & 1.140
        & 1.1
        \\

        2n, \ref{fig:kineticallyTrapped}c
        & 1f
        & Truncated cuboctahedron
        & $10^6$
        & 5
        & 0.6
        & 1.060
        & 1.105
        & 1.1
        \\

        2o, \ref{fig:kineticallyTrapped}d
        & 1k
        & Rhombicuboctahedron
        & $10^6$
        & 5
        & 0.6
        & 1.065
        & 1.115
        & 1.1
        \\

        2p
        & -
        & Octahedron
        & $10^6$
        & 5
        & 1
        & 0.6
        & 0.8
        & 1.2
        \\

        2q
        & -
        & Octahedron with rough facets
        & $10^6$
        & 2.5
        & 1
        & 0.6
        & 0.8
        & 1.2
        \\

        2r
        & -
        & \makecell[tl]{Octahedron with rough facets \\ and rounded tips and edges}
        & $10^6$
        & 1.25
        & 1
        & 0.6
        & 0.8
        & 1.2
        \\

        2s
        & -
        & (Almost a) sphere
        & $10^6$
        & $0.8\bar{3}$
        & 1
        & 0.6
        & 0.8
        & 1.2
        \\

        2t
        & -
        & Sphere with pseudo-dendrites
        & $10^6$
        & $0.\bar{3}$
        & 1
        & 0.6
        & 0.8
        & 1.2
        \\
        
        \bottomrule
    \end{tabular}
\end{table}

\begin{figure*}
    \centering
    \includegraphics[width=1.0\textwidth]{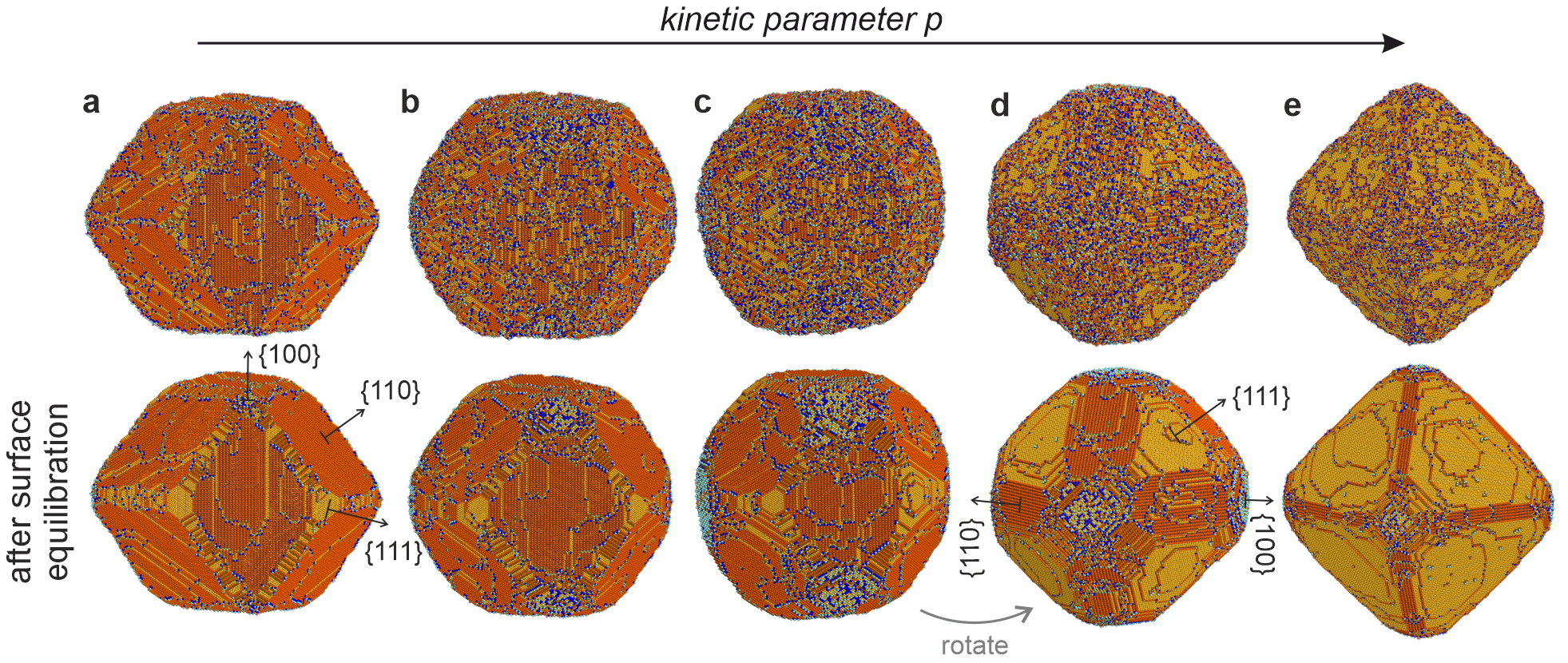}
    \caption{\textbf{Sensitivity of NC shape to the kinetic parameter $p$, that represents the proportion of growth events in rfKMC simulations.} Shape transformations between rhombic dodecahedron and octahedron. NCs grown to 1M atoms with $\epsilon^* = 5$, $\frac{E(z=3)}{E(z=5)} = 1.1$, $\frac{E(z=4)}{E(z=5)} = 1.3$, and (a)~$p=0.6$, (b)~$p=0.7$, (c)~$p=0.8$, (d)~$p=0.9$ and (e)~$p=1$ (irreversible growth).
    }
    \label{fig:SensitivityEquilibrium}
\end{figure*}

\begin{figure*}
    \centering
    \includegraphics[width=1.00\textwidth]{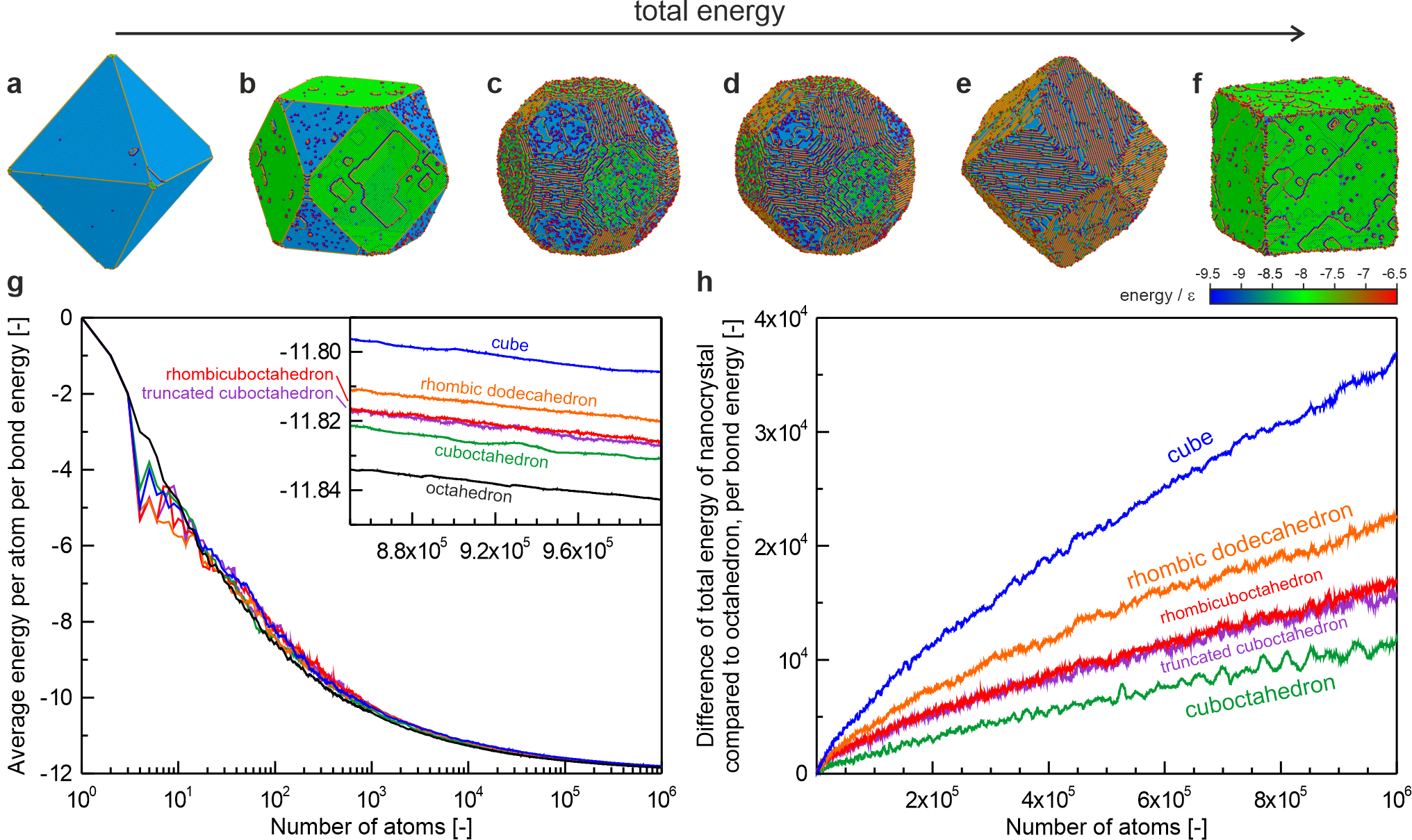}
    \caption{\textbf{Simulated NCs with shapes that are kinetically trapped in metastable equilibria.} Atoms are colored by energy. Shapes with increasing total energy: (a)~octahedron, (b)~cuboctahedron, (c)~truncated cuboctahedron, (d)~rhombicuboctahedron, (e)~rhombic dodecahedron, and (f)~cube. (g)~Variation of average energy per atom as the NC grows larger. It asymptotically tends towards the bulk energy ($-12$ when normalized by the bond energy $\epsilon$ in an fcc lattice), indicating a decreased significance of surface energy in larger NCs. (h)~Difference in the absolute energy for the various shapes compared to an octahedron (global energy minimum). Table~\ref{table:inputParameters} presents the simulation conditions.
    }
    \label{fig:kineticallyTrapped}
\end{figure*}

\begin{figure*}
    \centering
    \includegraphics[width=1.0\textwidth]{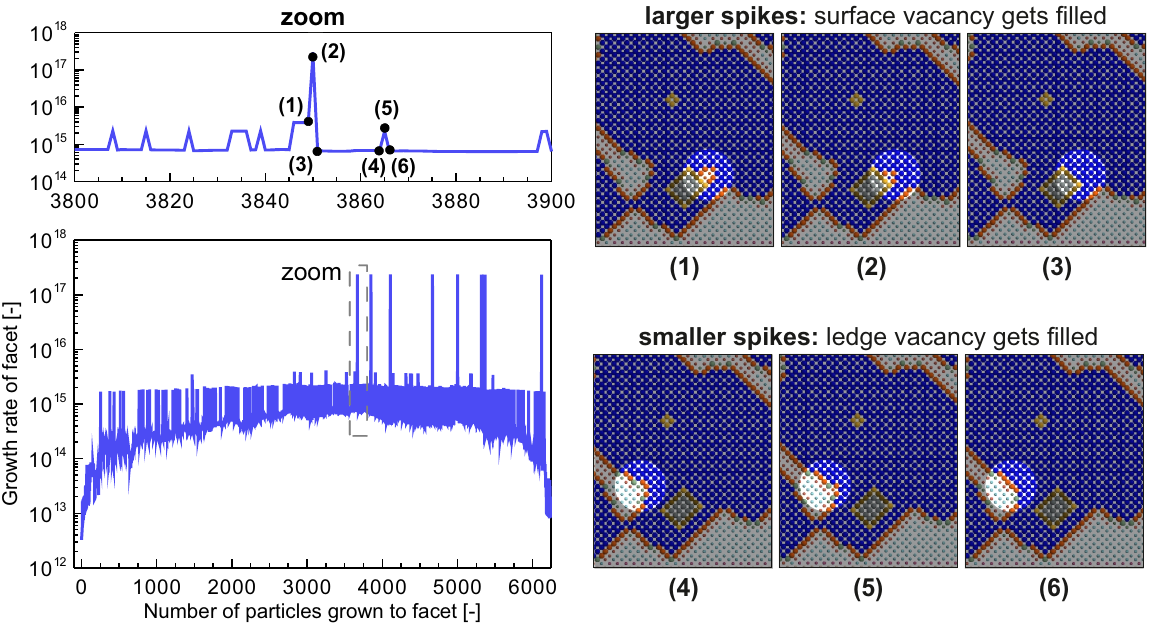}
    \caption{\textbf{Estimation of total growth rate on $\{100\}$ facets.} The spikes in the growth rate result from discrete states of growth rates (Boltzmann factors) as atoms incorporate into the lattice. The insets show that these spikes correspond to the filling of surface vacancies (1)-(3) and step vacancies (4)-(6). These spikes disappear upon integrating over the island size to estimate the growth time. Simulation parameters: $\epsilon^* = 5$, $p = 1$, $\frac{E(z=3)}{E(z=5)} = 0.6$ and $\frac{E(z=4)}{E(z=5)} = 0.8$.
    }
    \label{fig:Spikes}
\end{figure*}

\begin{figure}
    \centering
    \includegraphics[width=0.5\textwidth]{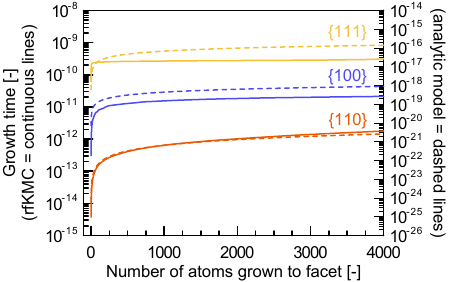}
    \caption{\textbf{Growth time to grow atoms to primary facets as estimated by the rfKMC simulations and analytic model.}
    }
    \label{fig:S5}
\end{figure}

\begin{figure*}
    \centering
    \includegraphics[width=1.0\textwidth]{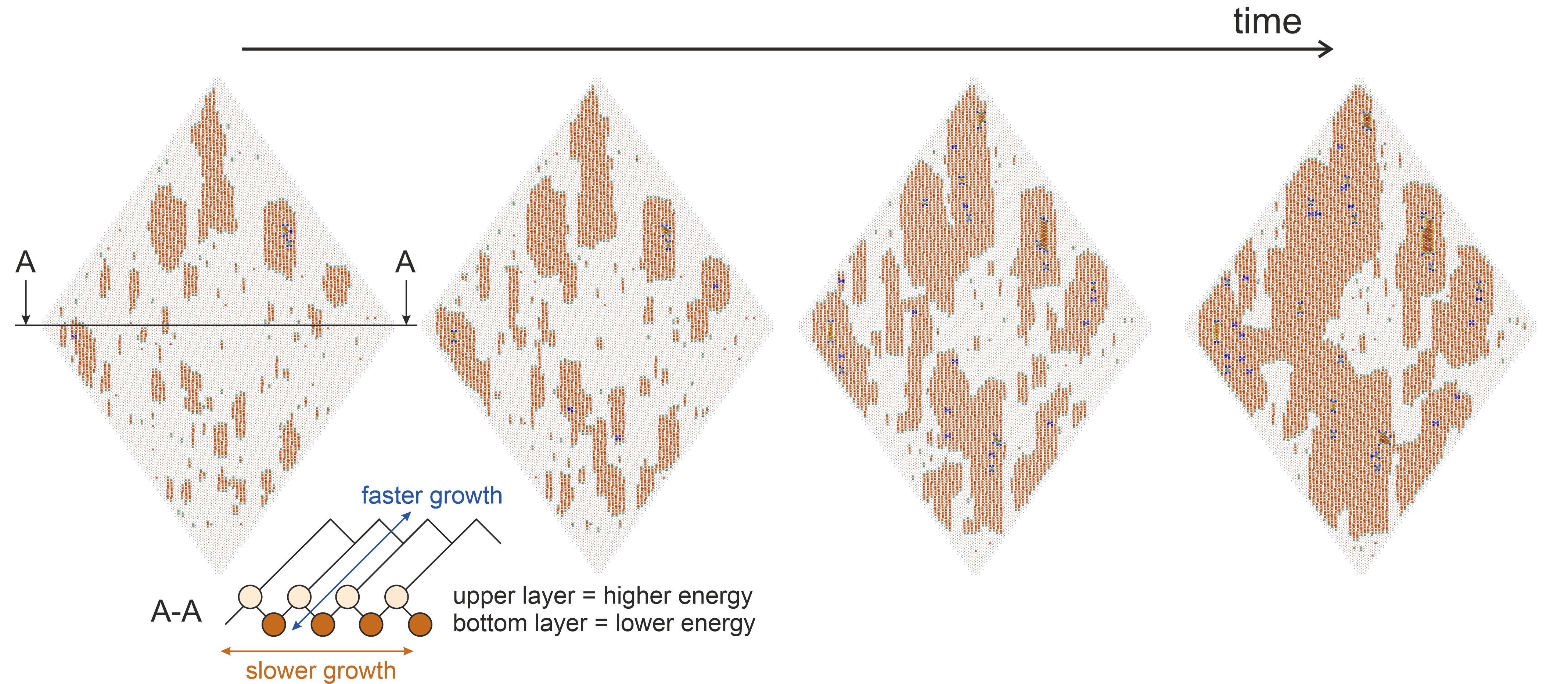}
    \caption{\textbf{Growth trajectory on $\{110\}$ facets of fcc NCs showing elongated (asymmetric) growth islands.} (Inset) The zigzag pattern on the $\{110\}$ facets creates energetic barriers that reduce growth velocity perpendicular to these facets, leading to elongated islands. Simulations for the growth of the first layer of atoms on a perfect $\{110\}$ facet, with $p = 1$, $kT = 0.04$~eV, and using the Lennard-Jones potential for silver with $r_0 = 0.2955$~nm, $\epsilon = 0.19774$~eV, and a cutoff radius of 1.2~nm.
    }
    \label{fig:elongated110Islands}
\end{figure*}

\begin{figure}
    \centering
    \includegraphics[width=1.0\textwidth]{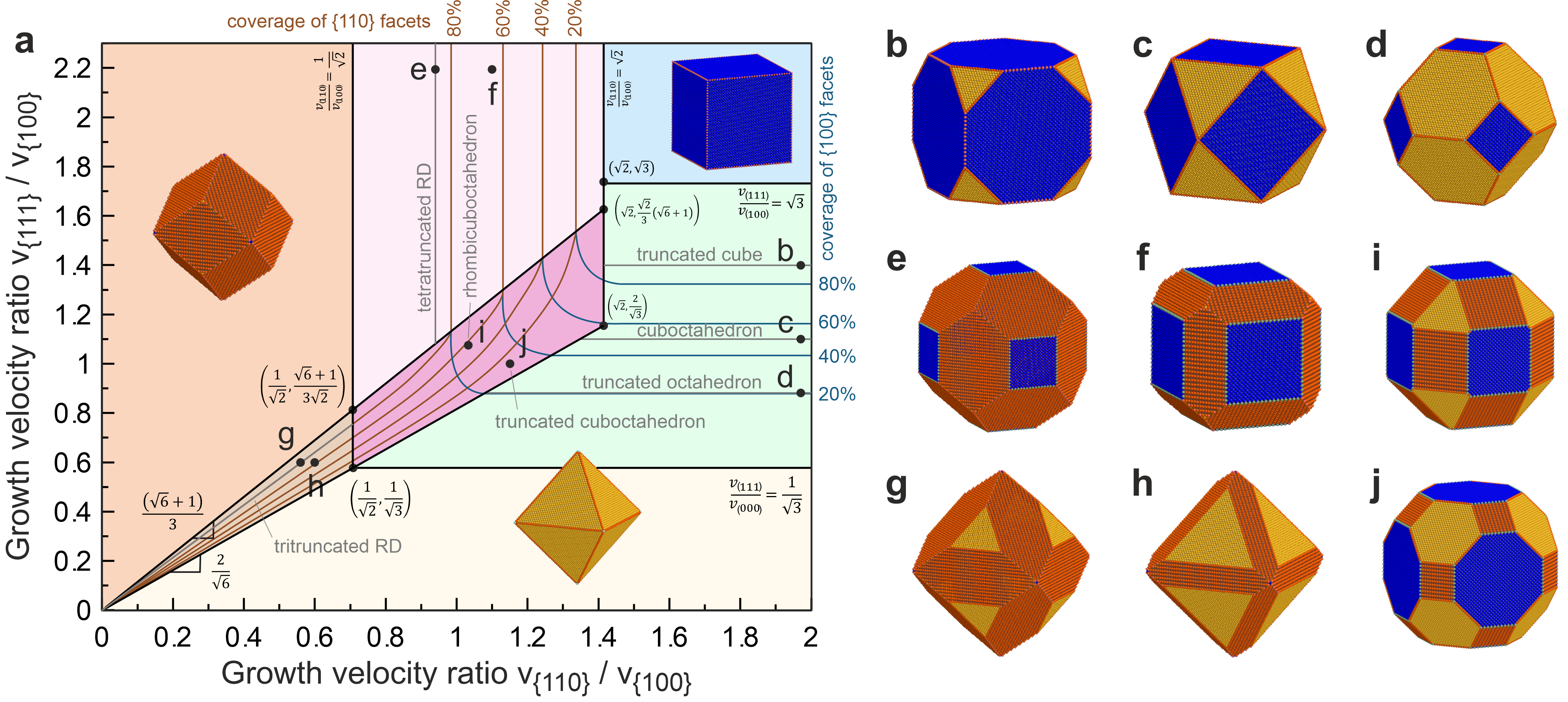}
    \caption{\textbf{Geometric construction of kinetic Wulff shapes.} (a)~Shape diagram based on growth velocities of primary facets $\{100\}$, $\{110\}$, and $\{111\}$. The diagram regions are color-coded to represent different shapes: cubes (blue), octahedra (yellow), and rhombic dodecahedra (orange), as well as truncations interpolating between (b-d)~cubes and octahedra (green), (e,f)~cubes and rhombic dodecahedra (pink), (g,h)~octahedra and rhombic dodecahedra (brown), and (i,j)~cubes, octahedra and rhombic dodecahedra (magenta). Equilateral truncated shapes include the truncated cube (b), the cuboctahedron (c), the truncated octahedron (d), the tetratruncated rhombic dodecahedron or equilateral chamfered cube (e), the tritruncated rhombic dodecahedra or equilateral chamfered octahedra (g), the rhombicuboctahedron (i), and the truncated cuboctahedron (j). The blue and brown isolines represent the percent of surface covered by $\{100\}$ and $\{110\}$ facets, respectively.
    }
    \label{fig:WulffConstruction}
\end{figure}

\begin{figure}
    \centering
    \includegraphics[width=1.0\textwidth]{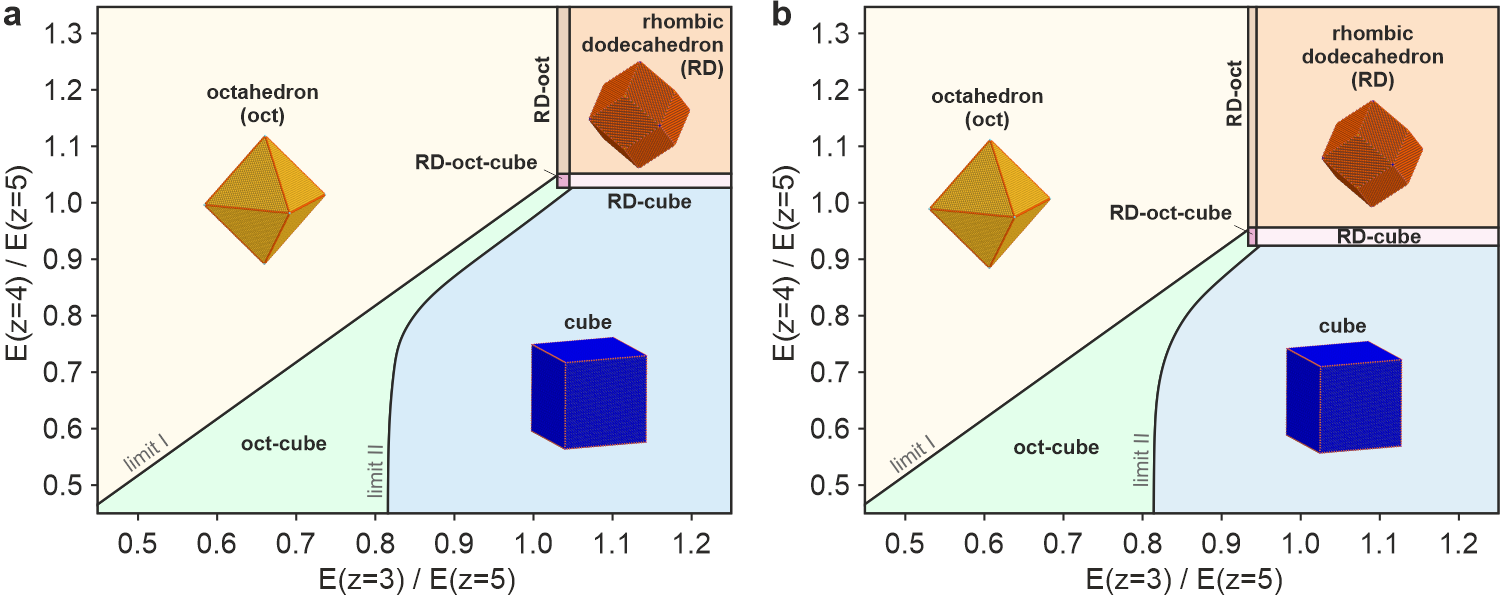}
    \caption{\textbf{Analytical shape diagram of fcc NCs showing the influence of energy associated with coordination number 6 in shape transition.} (a)~$\frac{E(z=6)}{E(z=5)} = 1.2$ and (b)~$\frac{E(z=6)}{E(z=5)} = 1.1$. Model estimated with $kT = 0.2$ for 3k atoms grown to each facet. Other energy levels depend linearly on coordination number, $E(z) = -z\epsilon$.
    }
    \label{fig:AnalyticModelEnergy6}
\end{figure}

\begin{figure*}
    \centering
    \includegraphics[width=1.0\textwidth]{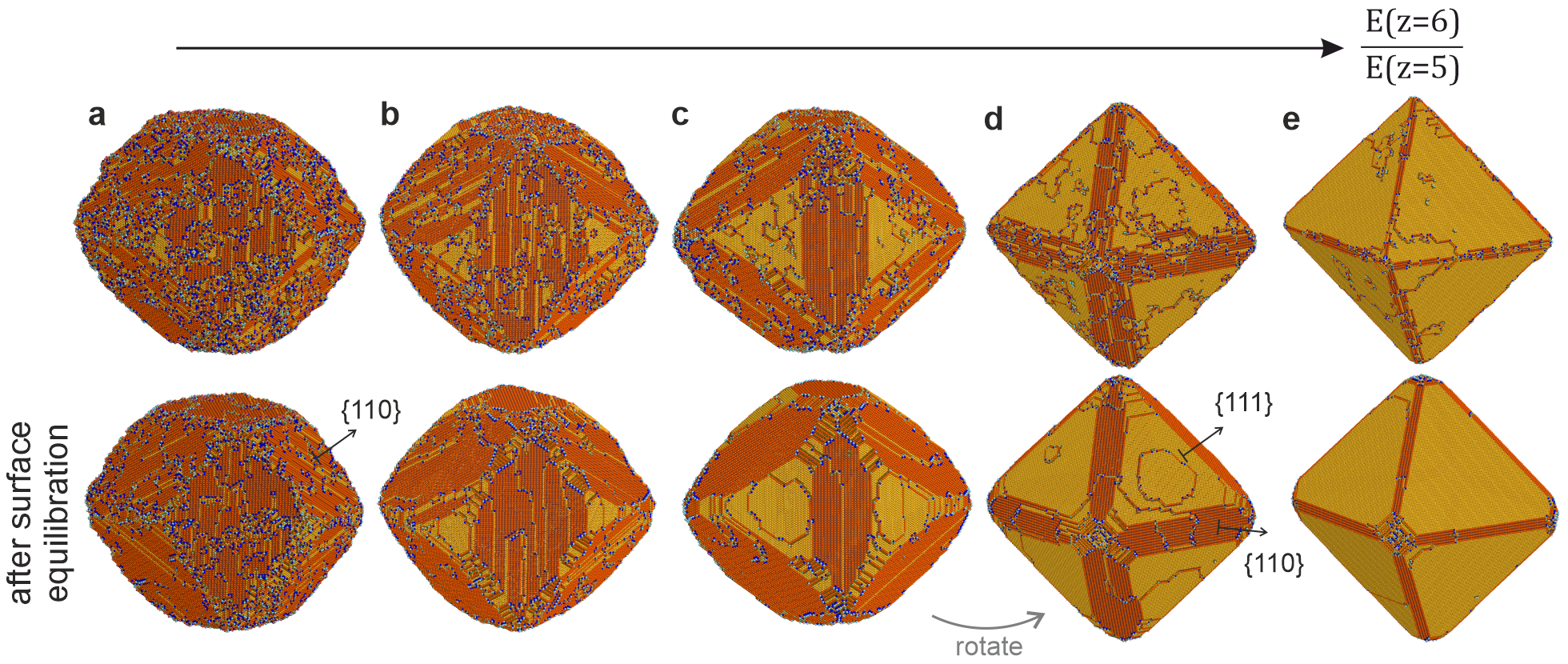}
    \caption{\textbf{Sensitivity of NC shape to the energy level of sites with coordination number 6.} Shape transformations between rhombic dodecahedra and octahedra. NCs grown to 1M atoms with $\epsilon^* = 5$, $\frac{E(z=3)}{E(z=5)} = 1.0$, $\frac{E(z=4)}{E(z=5)} = 1.3$, and varying $\frac{E(z=6)}{E(z=5)}$: (a) 1.00, (b) 1.12, (c) 1.14, (d) 1.16, and (e) 1.20.
    }
    \label{fig:SensitivityE6}
\end{figure*}

\begin{figure}
    \centering
    \includegraphics{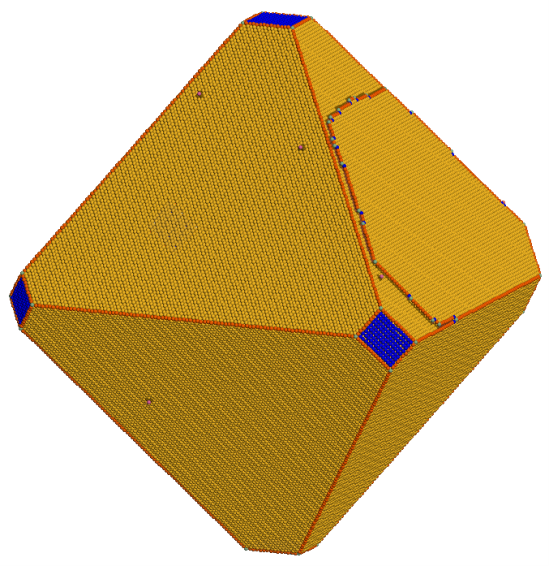}
    \caption{\textbf{NC shape for silver considering a Lennard-Jones potential.} NC grown to 1M atoms, with $p = 1$, $kT = 0.04$~eV, and Lennard-Jones parameters with a potential well of $\epsilon = 0.19774$~eV positioned at $r_0 = 0.2955$~nm, and considering a cutoff radius of 1.2~nm. The NC shape is an octahedron with truncations of $\{100\}$ tips to relieve energy (surface coverage of $\{100\}$ facets is $2.5\%$). The energy related to adatom nucleation at different crystallographic directions is $E^\text{ad}_{\{111\}} = -1.0169$~eV (coordination number 3), $E^\text{ad}_{\{100\}} = -1.1538$~eV (coordination number 4), and $E^\text{ad}_{\{110\}} = -1.3865$~eV (coordination number 5). Therefore, the energy ratios are $\frac{E(z=3)}{E(z=5)} \approx \frac{E^\text{ad}_{\{111\}}}{E^\text{ad}_{\{110\}}} = 0.73$ and $\frac{E(z=4)}{E(z=5)} \approx \frac{E^\text{ad}_{\{100\}}}{E^\text{ad}_{\{110\}}} = 0.83$. In the shape diagram of Fig.~2a,b of the manuscript, this results in an octahedron with minimal truncation of the tips, demonstrating the feasibility of predicting NC shapes from basic features of a potential energy model.
    }
    \label{fig:LennardJonesSilver}
\end{figure}

\begin{figure*}
    \centering
    \includegraphics[width=1.00\textwidth]{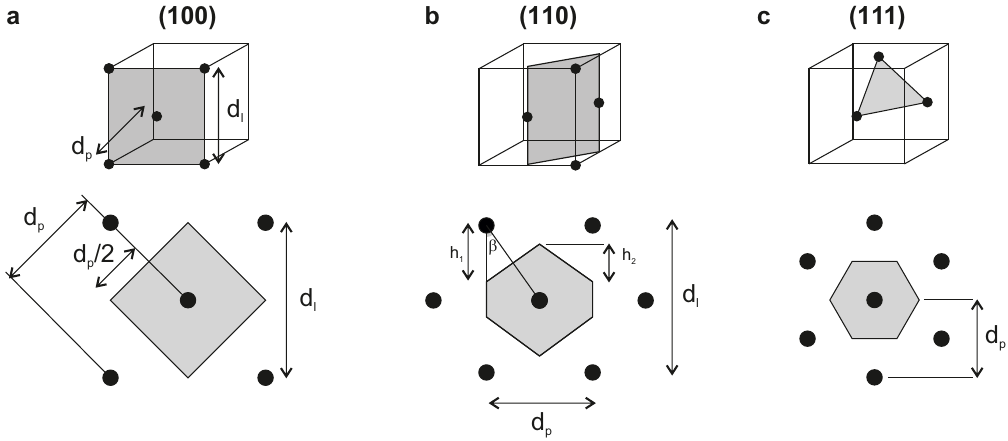}
    \caption{\textbf{Wigner-Seitz cell of an fcc lattice projected onto the primary facets to estimate surface area.} Table~\ref{table:WignerSeitzCell} provides values of surface area per atom.
    }
    \label{fig:WignerSeitzCell}
\end{figure*}

\begin{table}[h] 
    \centering
    \caption{\textbf{Surface area per atom on the primary fcc facets}, where $d_p$ is the interatomic distance, $d_l$ is the lattice parameter, and the heights $h_1$, $h_2$, and the angle $\beta$ are defined in Fig.~\ref{fig:WignerSeitzCell}.}
    \label{table:WignerSeitzCell}

    \begin{tabular}{ccc}
        \toprule
        $\{100\}$ & $\{110\}$ & $\{111\}$ \\
        \midrule
        $A_{\{100\}} = d_p^2$
        &
        $A_{\{110\}} = \left[ 2 \left( \frac{d_l}{2} - h_1 \right) + h_2 \right] $
        &
        $A_{\{111\}} = \frac{\sqrt{3}}{2} d_p^2$
        \\
        &       
        $h_1 = \frac{d_p}{4 \sin(\beta) \cos(\beta)}$
        &
        \\
        &
        $h_2 = \frac{d_p}{2} \tan(\beta)$
        &
        \\
        &
        $\beta = \arccot \left(\frac{d_l}{d_p} \right) = \arccot (\sqrt{2}) \approx 35.26 ^\circ$
        &
        \\
        \bottomrule
    \end{tabular}
\end{table}

\begin{figure*}
    \centering
    \includegraphics[width=1.00\textwidth]{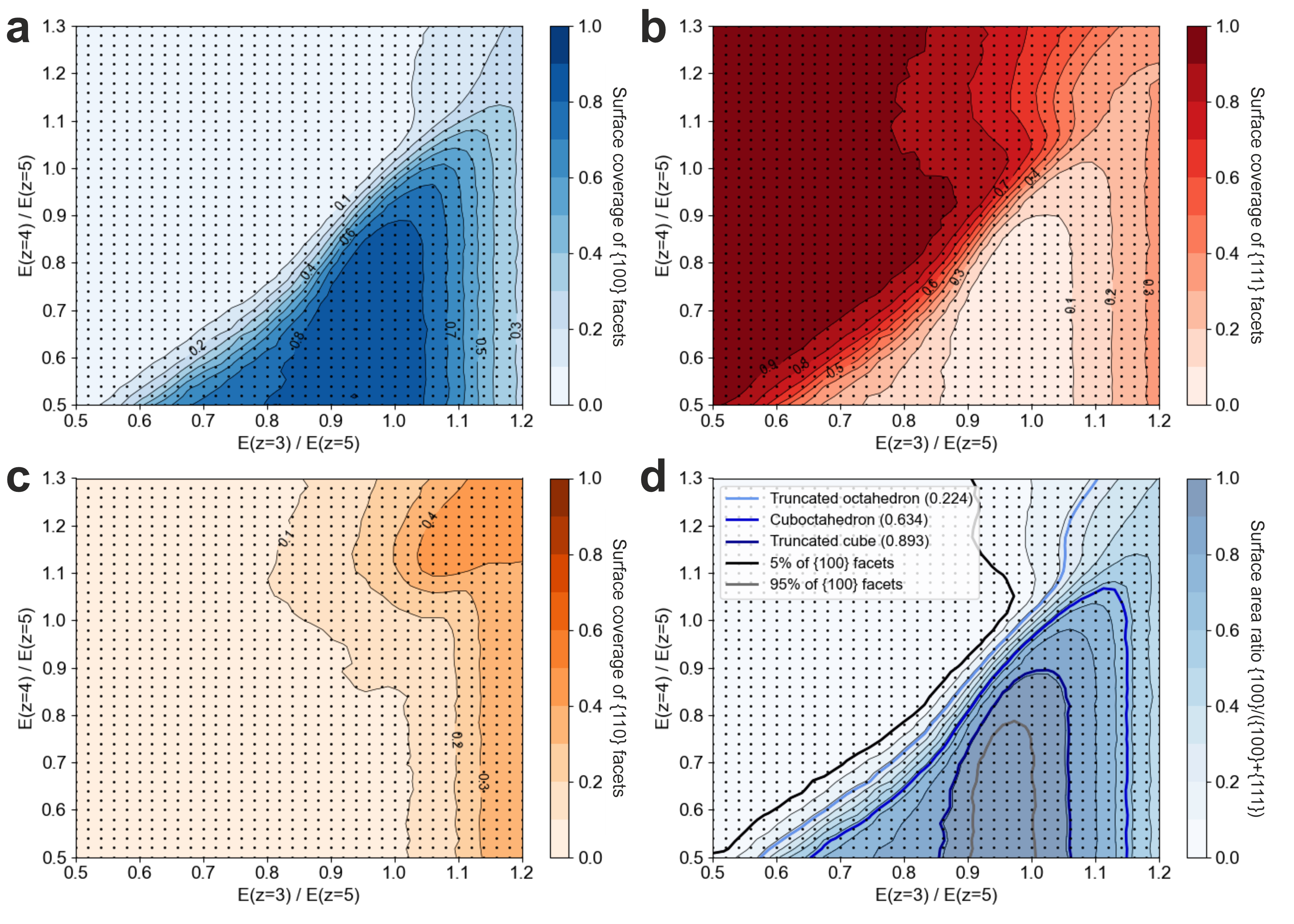}
    \caption{\textbf{Contour plots guiding the construction of the shape diagram in Fig.~2a of the manuscript.} Percentage of surface area covered by (a)~$\{100\}$, (b)~$\{111\}$, and (c)~$\{110\}$ facets. (d)~Ratio of surface area $r_A = \frac{A_{{100}}}{A_{{100}} + A_{{111}}}$. NCs grown to 1M atoms with $\epsilon^* = 5$ and $p = 1$.
    }
    \label{fig:ContourPlot_prop10}
\end{figure*}

\begin{figure*}
    \centering
    \includegraphics[width=1.00\textwidth]{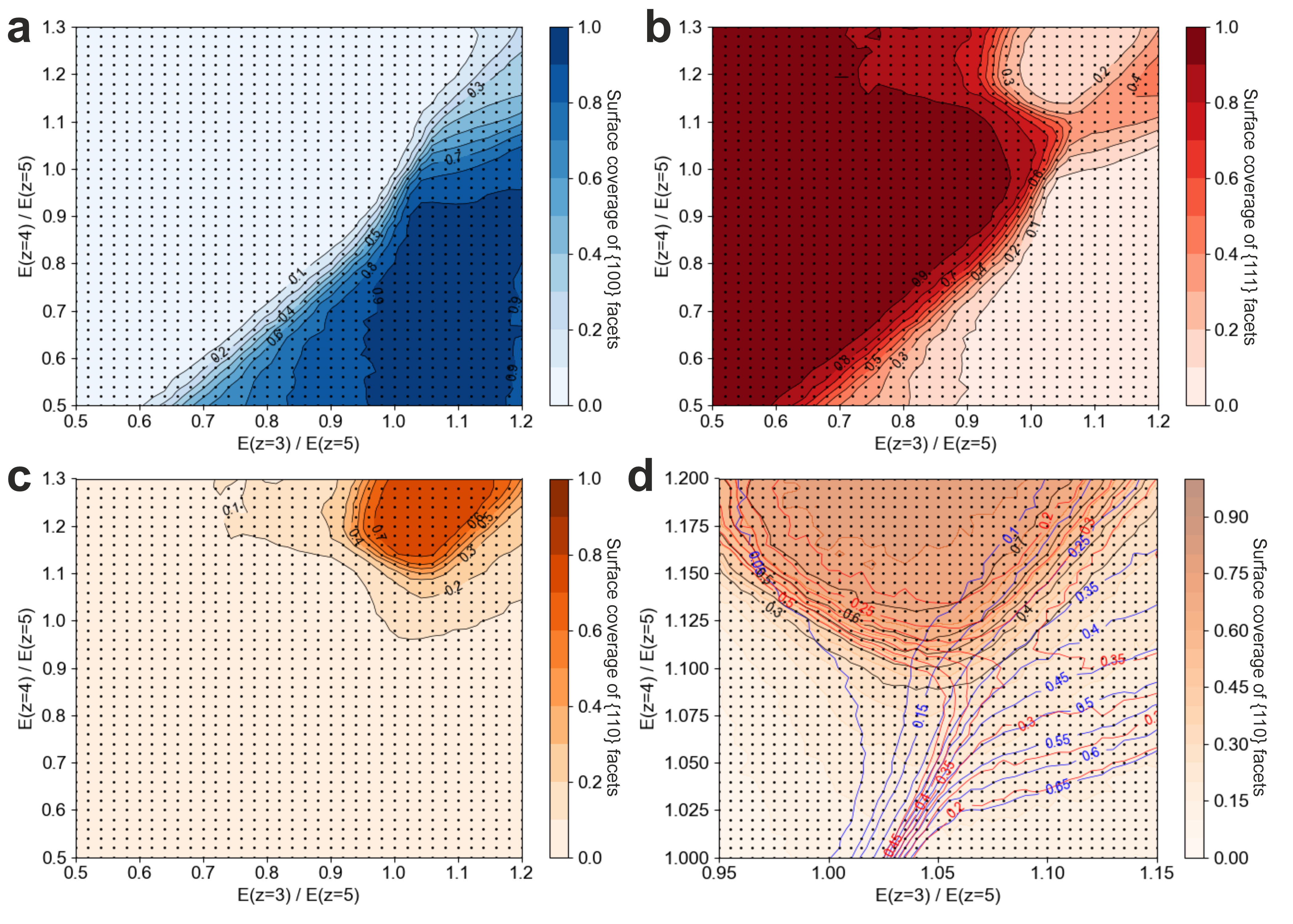}
    \caption{\textbf{Contour plots guiding the construction of the shape diagram in Fig.~2b-c.} Percentage of surface area covered by (a)~$\{100\}$, (b)~$\{111\}$, and (c)~$\{110\}$ facets. (d)~Zoomed-in view of the region where $\{110\}$ facets form, showing isolines of the three primary facets. NCs grown to 1M atoms with $\epsilon^* = 5$, $p = 0.6$, and $\frac{E(z=6)}{E(z=5)} = 1.1$.
    }
    \label{fig:ContourPlot_prop06_E6-55}
\end{figure*}

\begin{table}[h]
    \centering
    \caption{\textbf{Surface area of facets of equilateral shapes formed by more than one facet type.} The edge length of the solid is denoted by $a$. The surface coverage of the facet with Miller index \{hkl\} is defined as $\bar A_{\{hkl\}} = \frac{A_{\{hkl\}}}{A_{\{100\}} + A_{\{110\}} + A_{\{111\}}}$.}
    \label{table:ArchimedeanSolids}
    
    \begin{tabular}{ccccccc} 
        \toprule
        Shape
        & $A_{\{100\}}$
        & $A_{\{110\}}$
        & $A_{\{111\}}$
        & $\bar A_{\{100\}}$
        & $\bar A_{\{110\}}$
        & $\bar A_{\{111\}}$
        \\
        \midrule

        Truncated cube
        & $12 (1 + \sqrt{2}) a^2$
        & $0$
        & $2 \sqrt{3} a^2$
        & $\approx 0.893$
        & $0$
        & $\approx 0.107$
        \\

        Truncated octahedron
        & $6 a^2$
        & $0$
        & $12 \sqrt{3} a^2$
        & $\approx 0.224$
        & $0$
        & $\approx 0.776$
        \\

        Cuboctahedron
        & $6 a^2$
        & $0$
        & $2 \sqrt{3} a^2$
        & $\approx 0.634$
        & $0$
        & $\approx 0.366$
        \\

        Rhombicuboctahedron
        & $6 a^2$
        & $12 a^2$
        & $2 \sqrt{3} a^2$
        & $\approx 0.280$
        & $\approx 0.559$
        & $\approx 0.161$
        \\

        Truncated cuboctahedron
        & $12 (1 + \sqrt{2}) a^2$
        & $12 a^2$
        & $12 \sqrt{3} a^2$
        & $\approx 0.469$
        & $\approx 0.194$
        & $\approx 0.337$
        \\

        \makecell{Tritruncated \\ rhombic dodecahedron}
        & $0$
        & $(16 + \sqrt{2 + \sqrt{3}}) a^2$
        & $2 \sqrt{3} a^2$
        & $0$
        & $\approx 0.879$
        & $\approx 0.121$
        \\
        
        \makecell{Tetratruncated \\ rhombic dodecahedron}
        & $6 a^2$
        & $(16 + \sqrt{2 + \sqrt{3}} a^2$
        & $0$
        & $\approx 0.251$
        & $\approx 0.749$
        & $0$
        \\

        \bottomrule
    \end{tabular}
\end{table}

\begin{figure}
    \centering
    \includegraphics[scale = 0.95]{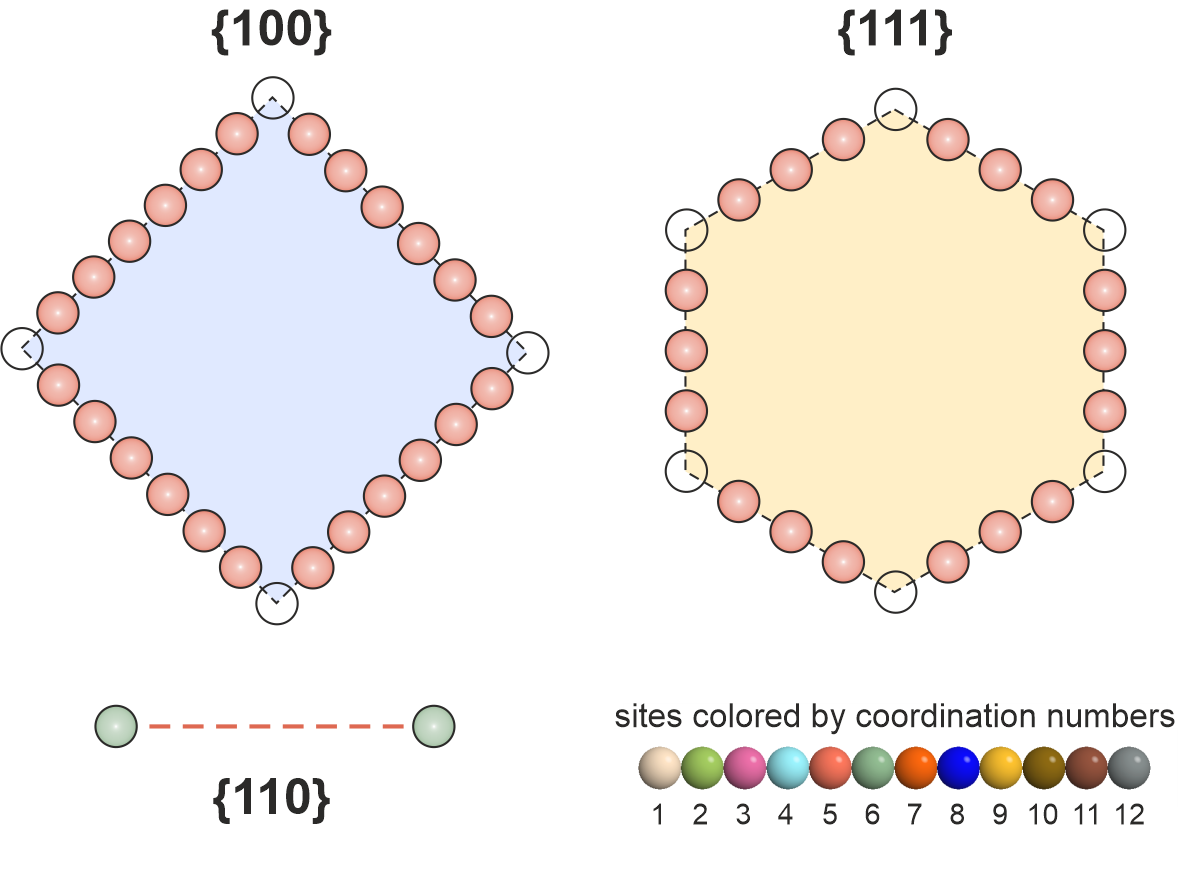}
    \caption{\textbf{Idealized geometry of growth islands.} The islands on the primary facets are squares on $\{100\}$ facets, lines on $\{110\}$, and regular hexagons on $\{111\}$. Growth sites at the corners (transparent) of the islands are neglected. The number of atoms $N$ in the growth island is treated as a continuous variable despite its discrete nature.} \label{fig:IslandsAnalyticModel}
\end{figure}